\documentclass[sigconf]{acmart}
\AtBeginDocument{%
  }

\copyrightyear{2025} 
\acmYear{2025} 
\setcopyright{cc}
\setcctype{by}
\acmConference[MM '25]{Proceedings of the 33rd ACM International Conference on Multimedia}{October 27--31, 2025}{Dublin, Ireland}
\acmBooktitle{Proceedings of the 33rd ACM International Conference on Multimedia (MM '25), October 27--31, 2025, Dublin, Ireland}
\acmDOI{10.1145/3746027.3754815}
\acmISBN{979-8-4007-2035-2/2025/10}

\usepackage{float}
\usepackage{placeins}
\usepackage{color}
\usepackage{amsmath}
\usepackage{xcolor}
\usepackage{xspace}
\usepackage{tcolorbox}
\usepackage{listings}
\usepackage{colortbl}
\usepackage{makecell}
\usepackage{booktabs}
\usepackage{multicol} %% 新增
\usepackage{multirow} %% 新增
\usepackage{subcaption}
\usepackage{graphicx}
\usepackage{adjustbox}
\usepackage{tcolorbox}

\makeatletter
\g@addto@macro\normalsize{%
  \abovedisplayskip 1pt plus2pt %minus1pt%
  \belowdisplayskip 
  \abovedisplayskip
  \abovedisplayshortskip  1pt plus3pt%
  \belowdisplayshortskip  1pt plus3pt% minus1pt%
}
\makeatother

\begin{document}
% \title{The Ghost in the Machine: Uncovering Source Bias in AI-Generated Videos}
\title{Generative Ghost: Investigating Ranking Bias Hidden in AI-Generated Videos}

% \affiliation{%
% \thanks{\ \ Corresponding author}
\author{Haowen Gao}
\email{gaohaowen23s@ict.ac.cn}
\orcid{0009-0007-9877-6163}
\affiliation{%
  \institution{State Key Lab of AI Safety, Institute of Computing Technology, CAS}
  \institution{University of Chinese Academy of Sciences}
  \city{Beijing}
  \country{China}
}

\author{Liang Pang}
\email{pangliang@ict.ac.cn}
\authornote{Corresponding author}
\affiliation{%
  \institution{State Key Lab of AI Safety, Institute of Computing Technology, CAS}
  \city{Beijing}
  \country{China}
}

\author{Shicheng Xu}
\email{xushicheng21s@ict.ac.cn}
\affiliation{%
  \institution{State Key Lab of AI Safety, Institute of Computing Technology, CAS}
  \institution{University of Chinese Academy of Sciences}
  \city{Beijing}
  \country{China}
}

\author{Leigang	Qu}
\email{leigangqu@gmail.com}
\affiliation{%
  \institution{National University of Singapore}
  \city{Singapore}
  \country{Singapore}
}

\author{Tat-Seng Chua}
\email{dcscts@nus.edu.sg}
\affiliation{%
  \institution{National University of Singapore}
  \city{Singapore}
  \country{Singapore}
}

\author{Huawei Shen}
\email{shenhuawei@ict.ac.cn}
\affiliation{%
  \institution{State Key Lab of AI Safety, Institute of Computing Technology, CAS}
  \city{Beijing}
  \country{China}
}

\author{Xueqi Cheng}
\email{cxq@ict.ac.cn}
\affiliation{%
  \institution{State Key Lab of AI Safety, Institute of Computing Technology, CAS}
  \city{Beijing}
  \country{China}
}

\renewcommand{\shortauthors}{Haowen Gao et al.}


\begin{abstract}

With the rapid development of AI-generated content (AIGC), the creation of high-quality AI-generated videos has become faster and easier, resulting in the Internet being flooded with all kinds of video content.
However, the impact of these videos on the content ecosystem remains largely unexplored. 
Video information retrieval remains a fundamental approach for accessing video content.
Building on the observation that retrieval models often favor AI-generated content in ad-hoc and image retrieval tasks, we investigate whether similar biases emerge in the context of challenging video retrieval, where temporal and visual factors may further influence model behavior.
To explore this, we first construct a comprehensive benchmark dataset containing both real and AI-generated videos, along with a set of fair and rigorous metrics to assess bias.
This benchmark consists of 13,000 videos generated by two state-of-the-art open-source video generation models. 
We meticulously design a suite of rigorous metrics to accurately measure this preference, accounting for potential biases arising from the limited frame rate and suboptimal quality of AIGC videos.
We then applied three off-the-shelf video retrieval models to perform retrieval tasks on this hybrid dataset. 
Our findings reveal a clear preference for AI-generated videos in retrieval.
Further investigation shows that incorporating AI-generated videos into the training set of retrieval models exacerbates this bias. 
Unlike the preference observed in image modalities, we find that video retrieval bias arises from both unseen visual and temporal information, making the root causes of video bias a complex interplay of these two factors. 
To mitigate this bias, we fine-tune the retrieval models using a contrastive learning approach. 
The results of this study highlight the potential implications of AI-generated videos on retrieval systems and offer valuable insights for future research in this area. 
Our dataset and code are publicly available at \url{https://github.com/Siaaaaaa1/video-source-bias}.

\end{abstract}

\begin{CCSXML}
<ccs2012>
   <concept>
       <concept_id>10002951.10003317</concept_id>
       <concept_desc>Information systems~Information retrieval</concept_desc>
       <concept_significance>500</concept_significance>
       </concept>
   <concept>
       <concept_id>10010147.10010178.10010224.10010225.10010231</concept_id>
       <concept_desc>Computing methodologies~Visual content-based indexing and retrieval</concept_desc>
       <concept_significance>500</concept_significance>
       </concept>
   <concept>
       <concept_id>10003120.10003121.10003122.10003334</concept_id>
       <concept_desc>Human-centered computing~User studies</concept_desc>
       <concept_significance>100</concept_significance>
       </concept>
   <concept>
       <concept_id>10010147.10010257.10010258.10010260</concept_id>
       <concept_desc>Computing methodologies~Unsupervised learning</concept_desc>
       <concept_significance>100</concept_significance>
       </concept>
 </ccs2012>
\end{CCSXML}

\ccsdesc[500]{Information systems~Information retrieval}
\ccsdesc[500]{Computing methodologies~Visual content-based indexing and retrieval}
\ccsdesc[100]{Human-centered computing~User studies}
\ccsdesc[100]{Computing methodologies~Unsupervised learning}

%\ccsdesc[500]{Information systems~Information retrieval}

\keywords{Text-Video Retrieval, AIGC, Bias and Fairness}



%\newpage
\maketitle

% \renewcommand{\thefootnote}{\fnsymbol{footnote}}
% \footnotetext[1]{\ Corresponding author}
\section{Introduction}

\begin{figure}
\centering
\includegraphics[width=0.9\linewidth]{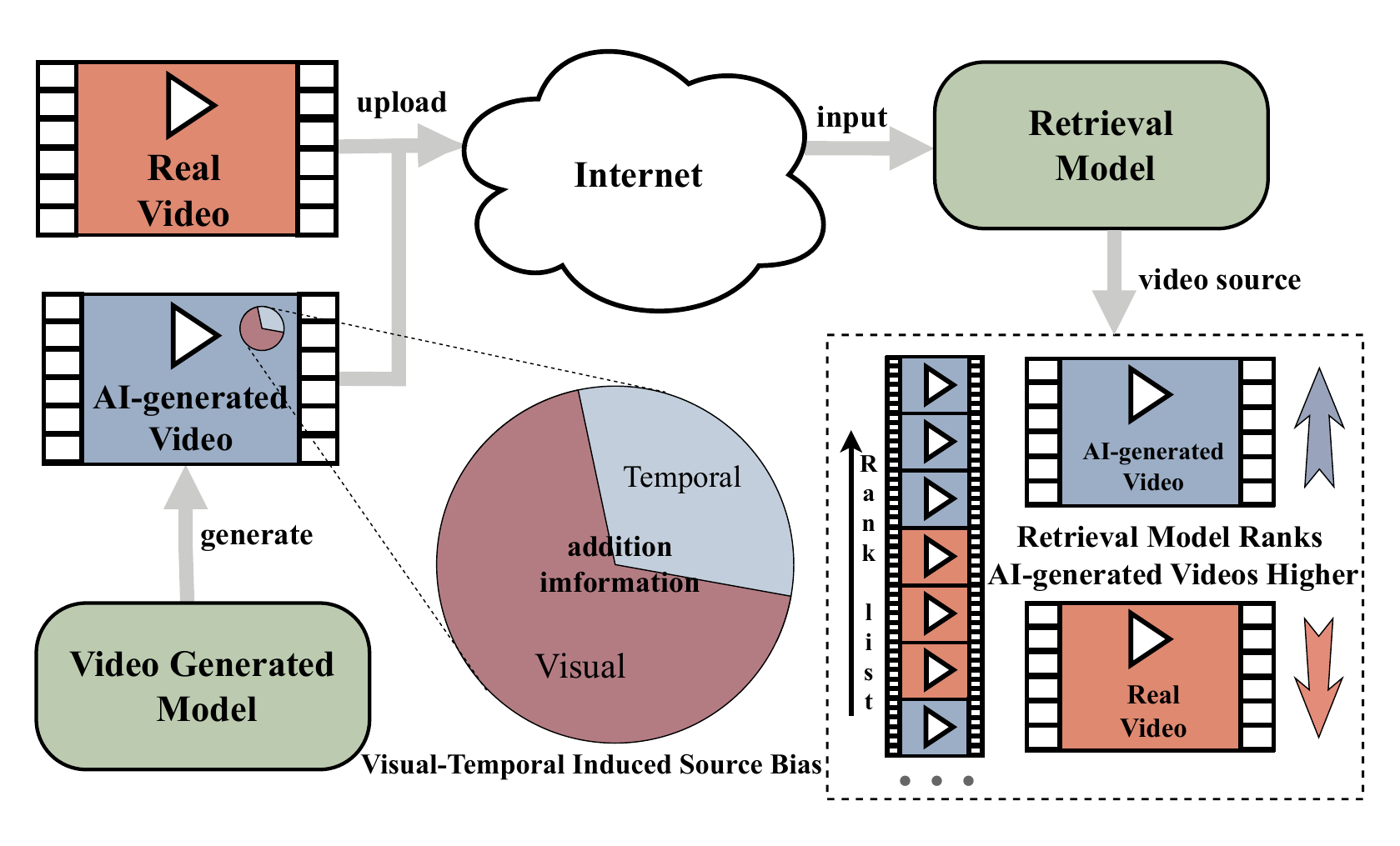}
\caption{The Visual-Temporal Induced Source Bias occurs when AI-generated videos, created by video generation models~\cite{yang2024cogvideox,opensora}, are mixed with real videos on the Internet. During text-video retrieval, the retrieval model~\cite{bain2021frozen,li2022align,wang2022internvideo} tends to prioritize AI-generated videos due to the extra visual and temporal information embedded by the generation model.}
\label{fig:overall}
\end{figure}

In the contemporary digital era, video content stands out among various media formats due to its unique dynamism and vividness, emerging as the preferred medium for information dissemination and entertainment~\cite{goggin2010global,tarchi2021learning}.
Information retrieval, particularly in the video domain, acts as the vital entry point for users navigating this vast content ecosystem.
As artificial intelligence (AI) rapidly evolves, the production of AI-generated videos has become significantly easier and faster~\cite{wu2023ai,foo2023ai,xing2024survey,opensora}, leading to a surge of this content type online. 
This influx of AI-generated videos raises a critical question: How will video retrieval models handle these AI-generated videos?

Similar questions have also been proposed in the textual and image content ecosystem~\cite{dai2023llms,xu2024invisible,zhou2024source}. 
Previous studies have found that in both text and image domains, retrieval models prioritize AI-generated content, a phenomenon known as "source bias"~\cite{dai2024neural,xu2023ai,dai2024bias}, the aim of our study is to explore the source bias in video modality.
However, the video modality presents unique challenges, making bias assessment more complex.
First, generating AI-generated videos that are semantically similar to real ones is particularly difficult due to the resource-intensive and time-consuming nature of video creation. This issue is further compounded by the limitations of open-source models, which often fail to produce satisfactory results.
Second, assessing bias in video retrieval models requires a more nuanced approach. It is necessary to incorporate multidimensional metrics across the retrieval list to capture various biases.
Additionally, the impact of semantic discrepancies between videos must be minimized to avoid skewing evaluation results. 
Specifically, it is crucial to ensure that the retrieval model does not favor AI-generated or real videos simply due to semantic proximity to the textual description.
Lastly, pinpointing the sources of bias is more challenging in video than in text or image domains. 
Videos contain not only rich visual information but also unique temporal elements, adding complexity to bias analysis.

The first two challenges have led us to focus on creating a standardized benchmark for video retrieval that includes both real and AI-generated videos, with relevance annotations for these videos also being developed. 
To the best of our knowledge, no such dataset currently exists.
To ensure that the benchmark is suitable for assessing video retrieval bias, we must address challenges related to video generation alignment (the similarity between queries and generated videos), video generation quality (the similarity between real and generated videos), and unbiased bias assessment metrics (which account for varying relevance levels) (See~\textbf{\color{red}{\S\ref{sec:2}}})~\cite{cocktail}.
Our dataset consists of 13,000 videos, including 9,000 training videos and four test sets, each containing 1,000 videos. It leverages two state-of-the-art video generation models: CogVideoX~\cite{yang2024cogvideox} and OpenSora V1.2~\cite{opensora}. These models collaboratively generate videos by integrating text, real video frames, or clips, all based on the MSR-VTT dataset~\cite{xu2016msr}.
To further minimize the impact of semantic similarity between videos and their corresponding queries on retrieval ranking, we introduce a novel evaluation metric, $Normalized\Delta$.
Additionally, to offer a more comprehensive assessment of multi-dimensional retrieval performance, we propose the MixR metric. 
Unlike previous evaluation measures that focus solely on recall@1 (R@1), MixR combines mean ranking (MeanR)  and median ranking (MedR) with R@1 to capture the broader impact of the entire retrieval list.

Experimental results from our constructed benchmark reveal an intriguing phenomenon: text-video retrieval models tend to prioritize AI-generated videos over real videos. 
Specifically, these models often rank AI-generated videos higher than real videos, even when both have the same relevance level (see \textbf{\color{red}{\S\ref{sec:3.2}}}). 
As AI-generated videos become more prevalent on the internet, they are likely to be incorporated into the training datasets of future retrieval models. 
Our further findings suggest that as the proportion of AI-generated videos in the training set increases, retrieval models progressively favor AI-generated content, exhibiting a growing bias toward prioritizing it (see \textbf{\color{red}{\S\ref{sec:3.3}}}).

Videos differ significantly from other modalities, as they contain several times more information than text and images.
Source bias in videos not only encompasses the visual bias found in image modalities but also integrates temporal information.
Compared to the singular causes of source bias in text and image modalities, the root causes of video source bias are far more complex and challenging to analyze.
In response to this challenge, we investigate the underlying causes of video source bias by disrupting temporal information through randomized frame order (see \textbf{\color{red}{\S\ref{sec:4.1}}}) and isolating visual information by extracting single-frame images (see \textbf{\color{red}{\S\ref{sec:4.2}}}).
Experiments demonstrate that the additional information embedded in both the visual and temporal components of generated videos plays a key role in generating bias, which we term \textbf{Visual-Temporal Induced Source Bias}. 
Specifically, real videos contain richer temporal information, whereas generated videos, lacking sufficient temporal depth, primarily rely on single-frame changes. 
This lack of depth contributes to the formation of source bias. 
Moreover, compared to retrieval tasks in other modalities, video retrieval exhibits a stronger preference for the first retrieved video. It shows a general tendency to favor AI-generated videos in the retrieval results.

To mitigate Visual-Temporal Induced Source Bias in retrieval models, we apply a contrastive learning approach~\cite{chen2020simple} to fine-tune the models, incorporating AI-generated videos into the training set (see \textbf{\color{red}{\S\ref{sec:5.1}}})~\cite{wang2025perplexity}.
Through fine-tuning, we train the model to prioritize real videos over AI-generated ones, placing real videos at the top of the retrieval list and ensuring they appear before AI-generated videos in the overall ranking, which effectively reduces Visual-Temporal Induced Source Bias.
By quantifying the differences in vector representations between the debiased and original models, we use t-SNE~\cite{van2008visualizing} to visualize the Visual-Temporal Induced Source Bias in the videos (See \textbf{\color{red}{\S\ref{sec:5.2}}}).

Our contributions include:

(1) We construct a benchmark that includes both real and AI-generated videos to investigate the impact of AI-generated content on video retrieval models. 

(2) We reveal that AI-generated videos introduce Visual-Temporal Induced Source Bias, which stems from the additional visual and temporal information embedded by video generation encoders, leading retrieval models to rank them higher.

(3) We propose a debiasing method for video retrieval models that effectively reduces Visual-Temporal Induced Source Bias towards AI-generated videos.

\section{Benchmark Construction}
\label{sec:2}
In this section, we pioneerly construct a benchmark to evaluate the impact of AI-generated videos on text-video retrieval models. The construction of this benchmark involves four stages: real retrieval dataset selection, semantic-equivalent video generation, dataset quality evaluation, and construction of bias evaluation metrics. Additionally, the benchmark should meet three key requirements to ensure the reliability of the research outcomes:

\textbf{(1) Identical Semantics:} Ensuring generated videos have the same semantics as original ones, so they share the same relevant labels as the query, which helps prevent abnormal retrieval rankings due to excessive video-query similarity.

\textbf{(2) Realistic Generation:} Video generation methods should align with real-world applications (e.g., generating videos from texts or combining texts with images) and ensure video quality, especially the similarity between real and generated videos.

\textbf{(3) Unbiased Assessment:} Ensuring identical semantics (the first requirement) is hard to precisely attain because of generation model capabilities. Therefore, we need robust metrics accounting for varying relevance levels to measure source bias impact and prevent the influence of different levels.

\begin{figure*}[h]
\centering

\begin{subfigure}{0.17\textwidth}
    \centering
    \includegraphics[width=\linewidth]{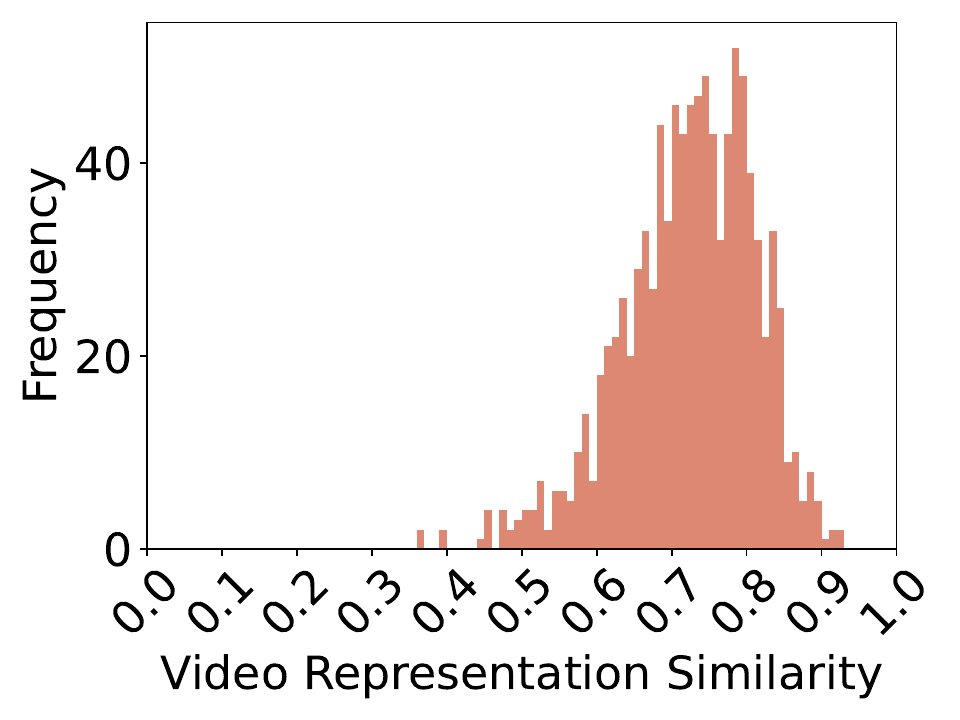}
    \caption*{\scriptsize(a) CogVideoX TextCond}
    \label{fig:REALR@k}
\end{subfigure}
\hfill
\begin{subfigure}{0.17\textwidth}
    \centering
    \includegraphics[width=\linewidth]{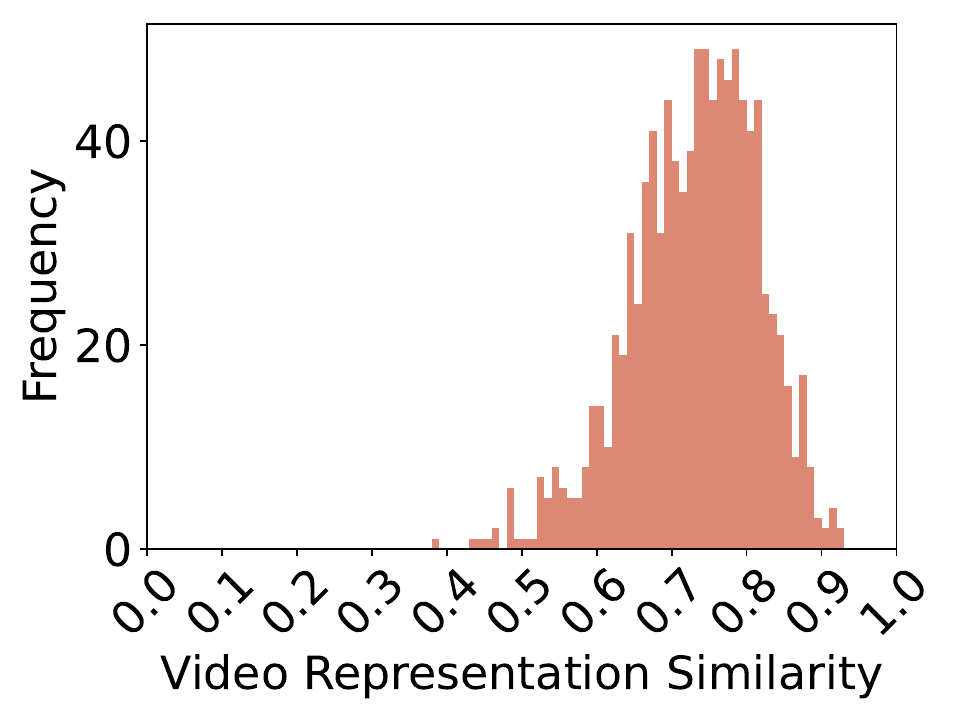}
    \caption*{\scriptsize(b) OpenSora TextCond}
    \label{fig:AIR@k}
\end{subfigure}
\hfill
\begin{subfigure}{0.17\textwidth}
    \centering
    \includegraphics[width=\linewidth]{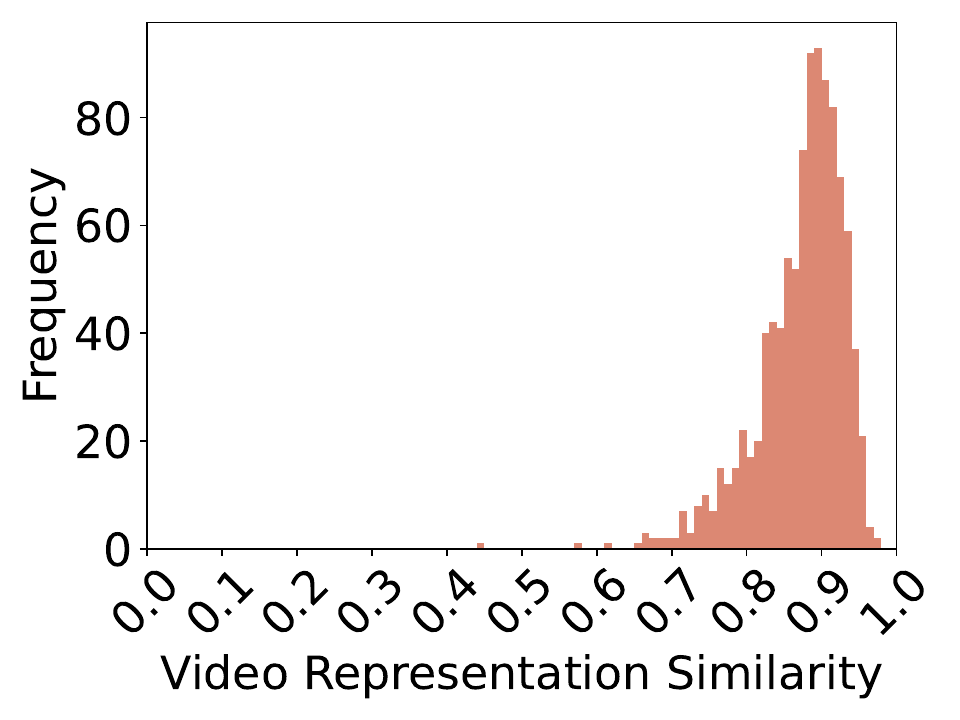}
    \caption*{\scriptsize(c) OpenSora ImageCond}
    \label{fig:mixRealR@k}
\end{subfigure}
\hfill
\begin{subfigure}{0.17\textwidth}
    \centering
    \includegraphics[width=\linewidth]{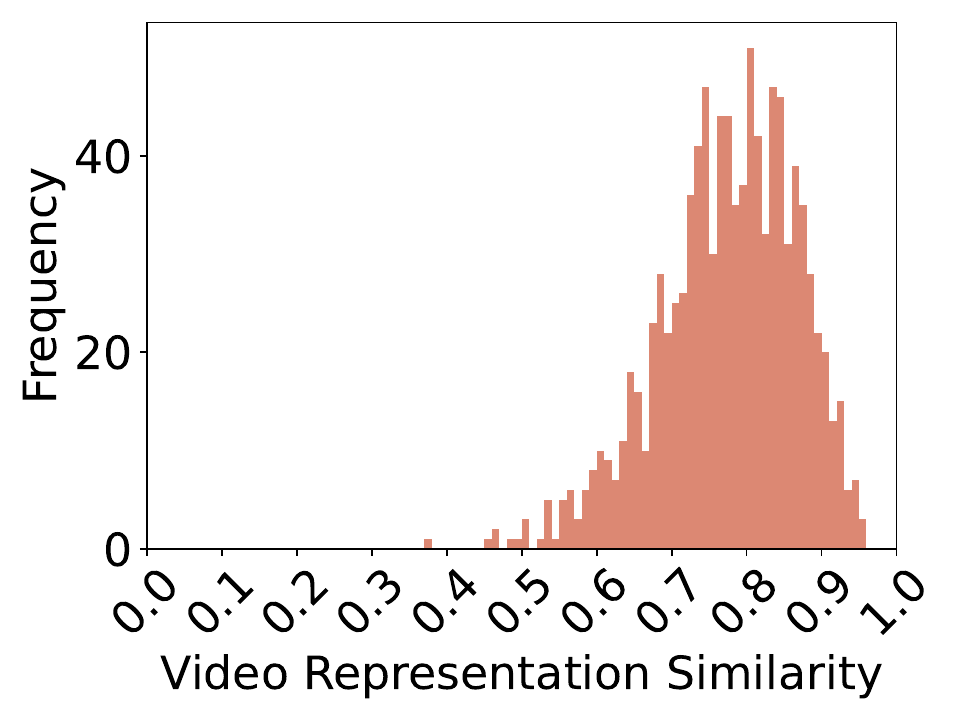}
    \caption*{\scriptsize(d) OpenSora VideoExt}
    \label{fig:mixAIR@k}
\end{subfigure}
\hfill
\begin{subfigure}{0.17\textwidth}
    \centering
    \includegraphics[width=\linewidth]{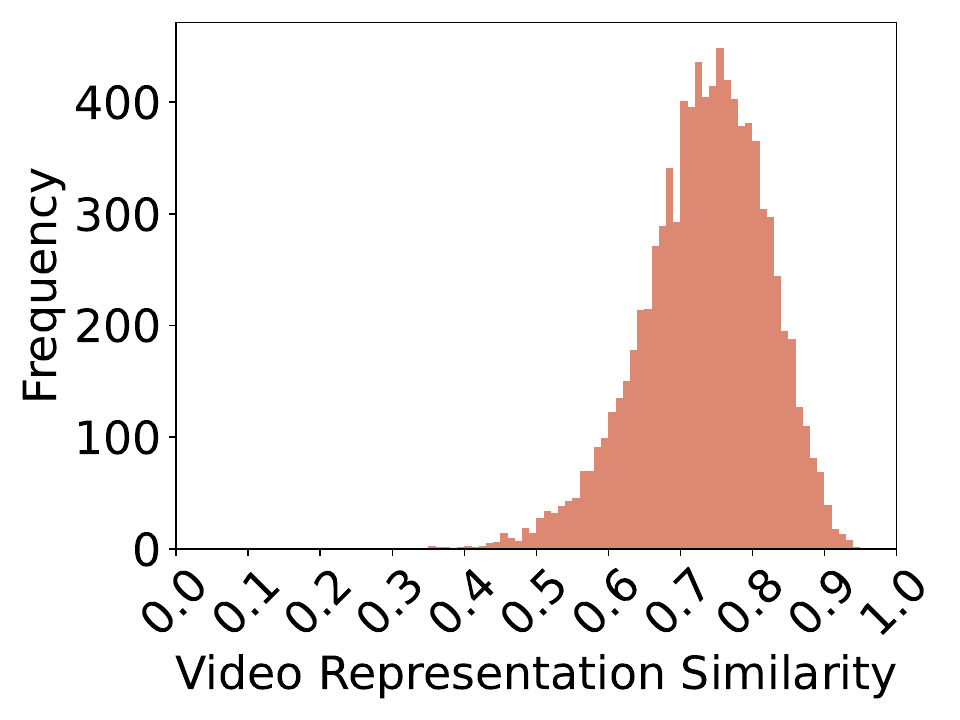}
    \caption*{\scriptsize(e) OpenSora TextCond(train)}
    \label{fig:relativedeltaR@k}
\end{subfigure}
\caption{Using the CLIP model to compute the similarity between AI-generated video datasets and real video datasets (X-axis), and analyzing the distribution of similarity frequencies (Y-axis).}
\label{fig:benchmark}
\end{figure*}

\begin{table}
\belowrulesep=0pt
  \aboverulesep=0pt
  \caption{AI-generated video Datasets Overview.}
  \label{tab:benchmark}
  \setlength{\tabcolsep}{0.8mm}
  \renewcommand{\arraystretch}{1.25}
  \adjustbox{max width=0.4\textwidth}{%
  \centering
  \begin{tabular}{lccccc}
  \toprule
    \makecell{Dataset} & \makecell{Number} & \makecell{FPS} & \makecell{Duration} & \makecell{Resolution} & \makecell{Similarity} \\
    \midrule
    \small CogVideoX TextCond & 1000 & 8 & 6.12 & 720×480 & 0.7232 \\
    \small OpenSora TextCond & 1000 & 24 & 4.25 & 640×360 & 0.7304 \\
    \small OpenSora ImageCond & 1000 & 24 & 4.25 & 640×360 & 0.8725 \\ 
    \small OpenSora VideoExt & 1000 & 12* & 8.50 & 424×240* & 0.7745 \\
    \small OpenSora TextCond (Train) & 9000 & 24 & 4.25 & 640×360 & 0.7332\\ 
    \bottomrule 
  \end{tabular}
  }
  \begin{minipage}{\textwidth}
  \footnotesize
  \raggedright
  \textit{\\ * Due to the limitation of resources.}
  \end{minipage}
\end{table}

\subsection{Real Retrieval Dataset Selection}
\label{sec:2.1}

To select an appropriate real retrieval dataset, we analyze five well-known text-video retrieval datasets, including MSR-VTT~\cite{xu2016msr}, MSVD~\cite{chen2011collecting}, DiDeMo~\cite{anne2017localizing}, ActivityNet~\cite{caba2015activitynet}, and LSMDC~\cite{rohrbach2015dataset}. The selected datasets should meet the following two core criteria. 
(1)~\textbf{Scenario Diversity}: they should include a diverse range of real videos, enriching the variety of video types encountered by users. This ensures reliability and broad applicability in evaluating source bias. For example, the LSMDC dataset does not meet this requirement.
(2)~\textbf{Annotation Completeness}: the captions for videos should cover the entire video, not just segments, and as many captions as possible should be provided. 
Without comprehensive annotations, accurate video generation cannot be achieved to minimize semantic biases between AI-generated and real videos. 
For instance, the MSVD, DiDeMo, and ActivityNet datasets do not meet this requirement.
Consequently, we select MSR-VTT, a widely recognized, large-scale dataset comprising 10,000 videos across 20 categories, with each video annotated by 20 English captions. 
The training set contains 9,000 videos, while the test set includes 1,000 videos. The dataset split follows the same partitioning as in Bain \textit{et al.}~\cite{bain2021frozen}.

\subsection{Semantic-equivalent Video Generation}
\label{sec:2.2}
Selecting appropriate video generation models and strategies is essential for generating semantically identical AI-generated videos for the MSR-VTT dataset. CogVideoX~\cite{yang2024cogvideox} and OpenSora V1.2~\cite{opensora} are two publicly available and widely used video generation models. They have distinct technical advantages, which make them suitable for comprehensively assessing the presence of source bias.

CogVideoX can generate high-quality content. However, it is time-consuming and has a single text-input interface. In contrast, OpenSora consumes fewer resources and has a more diverse interface, allowing the combination of images or videos for video generation. 
Given these state-of-the-art video generation models, generating semantically identical videos to the original ones remains a substantial challenge. Therefore, we employ multiple strategies to ensure comprehensive and robust experimental results.

For the videos in the test set, taking advantage of the diverse interfaces of OpenSora, we can adopt three strategies: text-only, text-image integration, and text-video integration for video generation. CogVideoX, on the other hand, can only use the text-only interface to generate videos. These four settings in the test set can more precisely verify the source bias.
For the videos in the training set, we simply use the highly efficient OpenSora with only text as the prompt input. This is to further explore the influence of AIGC in the training of the retrieval model. 
The detailed settings of these datasets are listed below:

\textbf{(1) Text-condition (TextCond):} To ensure that the generated videos are semantically similar to the original ones and encapsulate all relevant information, we integrate multiple captions into a single prompt.
For each video, we use GPT-4 to integrate its 20 captions by inputting them with the prompt: "I will provide 20 captions of the same video. Please assist in merging them into a comprehensive description." Through manual verification, the fusion caption is ensured to both summarize the multiple descriptions of the video and concisely represent the complete content of the actual video. Subsequently, we input the fusion caption into CogVideoX and OpenSora to obtain the CogVideoX TextCond and OpenSora TextCond (test and train) datasets.

\textbf{(2) Image-condition (ImageCond):} To enhance the alignment between generated video content and the visual semantics of real videos, we adopt a text-image integration approach. Specifically, we experiment with using keyframes extracted from the 0\%, 20\%, 40\%, 60\%, 80\%, and 100\% positions of real videos to guide video generation with OpenSora. Among these, keyframes from the 20\% position yield the best performance and are therefore selected to construct the OpenSora ImageCond dataset. Due to the high computational cost of video generation, this dataset includes only the test set.

\textbf{(3) Video-extending (VideoExt):} To capture the full content of a real video, we employ a text-video integration approach for video generation. Based on the OpenSora model's ability to use the earlier part of a video to supplement the later segments, we input the first half of a real video along with its fusion caption into OpenSora to create the OpenSora VideoExt dataset. This approach enhances the model's understanding of real videos, thereby increasing the likelihood of generating videos that are both visually and semantically consistent. Due to the high generation costs, the dataset contains only the test set.

\subsection{Dataset Quality Evaluations}
\label{sec:2.3}
To validate the reliability of the constructed benchmark, we compile key properties of the videos and assess their similarity to real videos. Key parameters of the dataset are shown in Table~\ref{tab:benchmark}. For similarity assessment, we uniformly select 10 frames from each generated video and its corresponding real video. Utilizing the CLIP ViT-B/32 model~\cite{radford2021learning}, we compute the average video representation and subsequently calculate the cosine similarity between the real and generated videos.

% We found that the AI-generated videos in the test set exhibit high semantic similarity to the real videos, with an average similarity exceeding 0.72, while the OpenSora ImageCond dataset achieves a similarity of 0.87. Additionally, the consistency between the training and corresponding test sets is well preserved. In the OpenSora TextCond dataset, the similarity between the train and test sets is nearly identical, with an average difference of a mere 0.0028.

We find that AI-generated videos in the test set exhibit a high semantic similarity to the real videos. Specifically, the average similarity exceeds 0.72, while the OpenSora ImageCond dataset achieves a similarity of 0.87. Additionally, the consistency between the training and corresponding test sets is well preserved. In the OpenSora TextCond dataset, the similarity between the training and test sets is nearly identical, with an average difference of only 0.0028.

Table \ref{tab:human} shows that on our benchmark, human evaluators determine that most real videos exhibit higher semantic similarity to the query and superior video quality compared to AI-generated videos, further ensuring the benchmark's reliability. Specifically, we recruit three evaluators with master's degrees to participate in the assessment. For each dataset, 100 groups are randomly sampled, with each group comprising a real video, a corresponding AI-generated video, and a query. The evaluators then select the video that demonstrates greater semantic similarity to the query and better overall video quality, and the selection ratios are subsequently tabulated.

\begin{table}
\belowrulesep=0pt
  \aboverulesep=0pt
  \caption{Human Evaluation of Video–Query Relevance}
  \label{tab:human}
  \centering
  % \setlength{\tabcolsep}{0.8mm}
  % \renewcommand{\arraystretch}{0.7}
  % \rowcolors{2}{white}{gray!15}
  \tiny
  \resizebox{0.35\textwidth}{!}{%
  \begin{tabular}{lccc}
  \toprule
    \makecell{Dataset} & Real & AI & Equal\\
    \midrule
    CogVideoX TextCond & 32\% & 15\% & 53\% \\
    OpenSora TextCond & 34\% & 18\% & 48\% \\
    OpenSora ImageCond & 47\% & 13\% & 40\% \\ 
    OpenSora VideoExt & 39\% & 14\% & 47\% \\
    \bottomrule 
  \end{tabular}
  }
  \footnotesize
\end{table}

\subsection{Bias Evaluation Metrics}
\label{sec:2.4}

% To offer a more fair and comprehensive evaluation of source bias, we propose two metrics: MixR and $Normalized\Delta$. Specifically, the MixR metric integrates the assessment of both the top-ranked retrieval and the entire retrieval list, while $\text{Normalized}\Delta$ mitigates the influence of semantic discrepancies in videos.

% \textbf{Notation:} Formally, for a single dataset, $REAL$ and $AI$ denote retrieval from the real and AI-generated video datasets, respectively. For a mixed dataset, $mixed-REAL$ and $mixed-AI$ correspond to retrieval tasks for real and AI-generated videos. $Metric$ encompasses R@k, MeanR, and MedR. Specifically, R@k measures the proportion of relevant items retrieved within the top-k results, MeanR calculates the average rank of relevant items, and MedR represents the median rank of relevant items across all queries. $Rank$ denotes the retrieval rank of a relevant item in a single query.

To fairly and comprehensively evaluate source bias in retrieval models, we introduce two metrics, $\text{MixR}$ and $\text{Normalized}\Delta$. In essence, the MixR metric aggregates performance over both the top-ranked retrieval results and the entire retrieval list, while $\text{Normalized}\Delta$ mitigates the influence of semantic discrepancies between videos.

\textbf{Notation:} Formally, for a single dataset, $\text{REAL}$ and $\text{AI}$ denote retrieval from the real and AI-generated video datasets. For a mixed dataset, $\text{mixed-REAL}$ and $\text{mixed-AI}$ correspond to retrieval tasks for real and AI-generated videos. $\text{Metric}$ encompasses $\text{R@k}$, $\text{MeanR}$, and $\text{MedR}$. Specifically, $\text{R@k}$ measures the proportion of relevant items retrieved within the top-k results, $\text{MeanR}$ calculates the average rank of relevant items, and $\text{MedR}$ represents the median rank of relevant items across all queries. $\text{Rank}$ denotes the retrieval rank of a relevant item in a single query.

\textbf{Basic Evaluation Metrics:} To quantify the source bias in mixed retrieval scenarios, we adopt the $\text{Relative}\Delta$ metric ~\cite{xu2024invisible}, which comprehensively measures how the presence of AI-generated videos influences the ranking of real videos. Specifically, the metric is defined as follows. When $\text{Metric} = \text{R@k}$, we set $s=1$, and when $\text{Metric} \in \{\text{MeanR}, \text{MedR}\}$, we set $s=-1$:
\begin{equation}
\text{Relative}\Delta = \frac{2s(\text{Metric}_{\text{mixed-REAL}} - \text{Metric}_{\text{mixed-AI})}}
{\text{Metric}_{\text{mixed-REAL}} + \text{Metric}_{\text{mixed-AI}}} \times 100\% ,
\label{eq:relative_delta}
\end{equation}

\textbf{Debias Evaluation Metrics:} Due to the high cost and instability of current video generation models, generated video quality is often inconsistent. This can bias retrieval models during source bias evaluation, unfairly favoring certain videos. To reduce the gap between $\text{Metric}_\text{{REAL}}$ and $\text{Metric}_\text{{AI}}$, we introduce two metrics: $\text{Location}\Delta$ and $\text{Normalized}\Delta$.

$\text{Location}\Delta$ estimates retrieval scores by comparing the ranking positions of real and AI-generated videos separately, without relying on semantic similarity. It serves as a reliable baseline for bias evaluation. $\text{Normalized}\Delta$ combines $\text{Relative}\Delta$ with $\text{Location}\Delta$, offering a more robust and accurate bias measurement by reducing the impact of semantic differences.

\textbf{Metrics Caculation:} To compute $\text{Location}\Delta$, we first record each video's ranking when retrieved separately from the real and AI-generated datasets. Since these datasets are not mixed, the rankings avoid cross-dataset semantic interference. We then simulate a mixed retrieval by interleaving rankings using:
\begin{align}
\text{Rank}_{\text{mixed-REAL}}&=\ 2*\text{Rank}_{\text{REAL}}-c,\\
\text{Rank}_{\text{mixed-AI}}&=2\ *\ \text{Rank}_{\text{AI}}-(1-c).
\label{eq:rank}
\end{align}

where $c \in {0, 1}$ is chosen randomly. Based on these, we compute mixed metrics $\text{Metric}_\text{{mixed-REAL}}^L$ and $\text{Metric}_\text{{mixed-AI}}^L$, and define $\text{Location}\Delta$ similarly to equation \ref{eq:relative_delta}..

% \begin{equation}
% Location\Delta=\frac{2s(Metric_{mixed-REAL}^L-Metric_{mixed-AI}^L)}{(Metric_{mixed-REAL}^L+Metric_{mixed-AI}^L)}.
% \label{eq:location_delta}
% \end{equation}

$\text{Relative}\Delta$ reflects the actual retrieval model performance, and $\text{Location}\Delta$ accounts for performance without considering semantic differences between real and AI-generated videos. 
The difference between these metrics represents the Visual-Temporal Induced Source Bias in the mixed retrieval model, termed $\text{Normalized}\Delta$:
\begin{equation}
\text{Normalized}\Delta = \text{Relative}\Delta - \text{Location}\Delta.
\label{eq:normalized_delta}
\end{equation}

When $\text{Normalized}\Delta > 0$, the model favors real videos over AI-generated ones. 
When $\text{Normalized}\Delta < 0$, the model favors AI-generated videos over real ones. 
Compared to directly using $\text{Relative}\Delta$, $\text{Normalized}\Delta$ better mitigates the influence of semantic discrepancies, offering a more accurate bias assessment.

Additionally, $\text{MixR}$ metric provides a unified evaluation by aggregating performance across both the top-ranked and the full retrieval list. Specifically, $\text{MixR}$ integrates $\text{R@1}$, $\text{MedR}$, and $\text{MeanR}$ using a combined score. Let $\Delta$ denote one of the variations: $\text{Relative}\Delta$, $\text{Location}\Delta$, or $\text{Normalized}\Delta$. The $\text{MixR}$ is then computed as:
\begin{equation}
\Delta \text{MixR} = (\Delta \text{R@1}+\Delta \text{MedR}+\Delta \text{MeanR}) / 3.
\label{eq:mixR}
\end{equation}

\begin{table*}[ht]
    \belowrulesep=0pt
    \aboverulesep=0pt
    \caption{The retrieval performance of different models is evaluated on four benchmarks we constructed. A positive $\text{Relative}\Delta$ or $\text{Normalized}\Delta$ indicates the model favors ranking real videos higher, while a negative \textcolor{red}{$\text{Relative}\Delta$} or \textcolor{red}{$\text{Normalized}\Delta$} suggests a preference for AI-generated videos. The absolute values of these metrics reflect the magnitude of bias. $\text{Normalized}\Delta$ includes a penalty term, offering a more accurate bias measurement than $\text{Relative}\Delta$.}
  \label{tab:main-exp-text}
  \centering
  \setlength{\tabcolsep}{0.8mm}
  \renewcommand{\arraystretch}{0.85}
  \small
  \resizebox{0.95\textwidth}{!}{%
  % \begin{tabular}{ll|cccc|cccc|cccc|cccc}
  \begin{tabular}{llcccccccccccccccc}
  \toprule
    \multicolumn{2}{c}{\makecell{\textbf{Dataset}}}  & 
    \multicolumn{4}{c}{\makecell{\textbf{CogVideoX TextCond}}} &
    \multicolumn{4}{c}{\makecell{\textbf{OpenSora TextCond}}} &
    \multicolumn{4}{c}{\makecell{\textbf{OpenSora ImageCond}}} &
    \multicolumn{4}{c}{\makecell{\textbf{OpenSora VideoExt}}}
    \\ \midrule
     Model & $\text{Metric}$ & $\text{R@1}$ & $\text{MedR}$ & $\text{MeanR}$ & $\text{MixR}$ & $\text{R@1}$ & $\text{MedR}$ & $\text{MeanR}$ & $\text{MixR}$ & $\text{R@1}$ & $\text{MedR}$ & $\text{MeanR}$ & $\text{MixR}$ & $\text{R@1}$ & $\text{MedR}$ & $\text{MeanR}$ & $\text{MixR}$\\ \midrule
     \multirow{6}{*}{\makecell[c]{Alpro}}& REAL & 24.10  & 8.00  & 49.61 & - & 24.10  & 8.00  & 49.61 & - & 24.1 & 8 & 49.61 & - & 24.1 & 8 & 49.61 & - \\
      & AI & 30.50  & 5.00  & 40.14 & - & 37.00 & 3.00  & 27.72 & - & 29.6 &  4 & 33.59 & - & 32.1 &  4 & 36.42 & -  \\
     & mixed-REAL & 10.10 & 14.00  & 82.94  & -  & 10.80 & 13.50  & 83.72  & -  & 8 & 15.5 & 94.31 & - & 8.7 & 17 & 95.90 & -  \\ 
     & mixed-AI & 22.60 & 10.00  & 101.16  & -  & 24.50 & 6.00  & 69.39  & - & 22.4 & 7 & 70.33 & - & 23.7 & 7 & 75.38 & -  \\ 
     & \cellcolor{gray!15}$\text{Relative}\Delta$ & \cellcolor{gray!15}\textcolor{red}{-76.45}  & \cellcolor{gray!15}\textcolor{red}{-33.33}  & \cellcolor{gray!15}19.80  & \cellcolor{gray!15}\textcolor{red}{-29.99}  & \cellcolor{gray!15}\textcolor{red}{-77.62}  & \cellcolor{gray!15}\textcolor{red}{-76.92}  & \cellcolor{gray!15}\textcolor{red}{-18.71}  & \cellcolor{gray!15}\textcolor{red}{-57.75} & \cellcolor{gray!15}\textcolor{red}{-94.74} & \cellcolor{gray!15}\textcolor{red}{-75.56} & \cellcolor{gray!15}\textcolor{red}{-29.13} & \cellcolor{gray!15}\textcolor{red}{-66.48} & \cellcolor{gray!15}\textcolor{red}{-92.59}  & \cellcolor{gray!15}\textcolor{red}{-83.33} & \cellcolor{gray!15}\textcolor{red}{-23.97} & \cellcolor{gray!15}\textcolor{red}{-66.63} \\ 
     & \cellcolor{gray!15}$\text{Normalized}\Delta$ & \cellcolor{gray!15}\textcolor{red}{-53.01}  & \cellcolor{gray!15}14.67  & \cellcolor{gray!15}41.02  & \cellcolor{gray!15}0.89  & \cellcolor{gray!15}\textcolor{red}{-35.39} & \cellcolor{gray!15}18.32  & \cellcolor{gray!15}38.26  & \cellcolor{gray!15}7.06 & \cellcolor{gray!15}\textcolor{red}{-74.26} & \cellcolor{gray!15}\textcolor{red}{-5.99} & \cellcolor{gray!15}9.61 & \cellcolor{gray!15}\textcolor{red}{-23.55} & \cellcolor{gray!15}\textcolor{red}{-64.12} & \cellcolor{gray!15}\textcolor{red}{-13.76} & \cellcolor{gray!15}6.87 & \cellcolor{gray!15}\textcolor{red}{-23.67} \\ 
     \multirow{6}{*}{\makecell[c]{Frozen}} & REAL & 22.90 & 8.00  & 49.81 & - & 22.90  & 8.00  & 49.81 & - & 22.9 & 8 & 49.81 & - & 22.9 & 8 & 49.811 & - \\
     & AI & 29.80 & 5.00  & 39.98 & - & 31.50  & 4.00  & 31.56 & - & 25.7 & 5 & 37.93 & - & 28.3 & 5 & 37.34 & -  \\
     & mixed-REAL & 6.90 & 20.00  & 92.25  & -  & 8.90  & 17.00  & 90.35  & - & 9.1 &  18 & 94.78 & - & 8.3 & 21 & 104.51 & -  \\ 
     & mixed-AI & 23.80  & 8.00  & 90.98  & -  & 25.50  & 7.00  & 72.41  & - & 18.9 & 9 & 80.01 & - & 21.6 & 8 & 71.89 & -  \\ 
     & \cellcolor{gray!15}$\text{Relative}\Delta$ & \cellcolor{gray!15}\textcolor{red}{-110.10}  & \cellcolor{gray!15}\textcolor{red}{-85.71}  & \cellcolor{gray!15}\textcolor{red}{-1.39}  & \cellcolor{gray!15}\textcolor{red}{-65.73}  & \cellcolor{gray!15}\textcolor{red}{-96.51} & \cellcolor{gray!15}\textcolor{red}{-83.33}  & \cellcolor{gray!15}\textcolor{red}{-22.05}  & \cellcolor{gray!15}\textcolor{red}{-67.30} &   \cellcolor{gray!15}\textcolor{red}{-70} & \cellcolor{gray!15}\textcolor{red}{-66.67} & \cellcolor{gray!15}\textcolor{red}{-16.9} & \cellcolor{gray!15}\textcolor{red}{-51.19} & \cellcolor{gray!15}\textcolor{red}{-88.96} & \cellcolor{gray!15}\textcolor{red}{-89.66} & \cellcolor{gray!15}\textcolor{red}{-36.99} & \cellcolor{gray!15}\textcolor{red}{-71.87} \\ 
     \multirow{-6}{*} & \cellcolor{gray!15}$\text{Normalized}\Delta$ & \cellcolor{gray!15}\textcolor{red}{-83.91} & \cellcolor{gray!15}\textcolor{red}{-37.71}  & \cellcolor{gray!15}20.63  & \cellcolor{gray!15}\textcolor{red}{-33.66}  & \cellcolor{gray!15}\textcolor{red}{-64.89} & \cellcolor{gray!15}\textcolor{red}{-13.76}  & \cellcolor{gray!15}23.08  & \cellcolor{gray!15}\textcolor{red}{-18.52} & \cellcolor{gray!15}\textcolor{red}{-58.48}  & \cellcolor{gray!15}\textcolor{red}{-18.67} & \cellcolor{gray!15}10.34 & \cellcolor{gray!15}\textcolor{red}{-22.27} & \cellcolor{gray!15}\textcolor{red}{-67.87}  & \cellcolor{gray!15}\textcolor{red}{-41.66} & \cellcolor{gray!15}\textcolor{red}{-8.21} & \cellcolor{gray!15}\textcolor{red}{-39.25} \\ 
     \multirow{6}{*}{\makecell[c]{Intern\\Video}} & REAL & 40.60 & 2.00  & 22.27 & - & 40.60  & 2.00  & 22.27 & - & 40.6 & 2 & 22.27 & - & 40.6 & 2 & 22.27 & -  \\
     & AI & 40.20  & 2.00  & 25.30 & - & 47.20  & 2.00  & 17.85 & - & 42.7 & 2 & 18.62 & - & 46.6 & 2 & 17.62 & - \\
     & mixed-REAL & 19.60  & 5.00  & 43.39  & -  & 27.40  & 5.00  & 74.16  & - & 29.1 & 4 & 83.65 & - & 28.2 & 4 & 75.72 & -  \\ 
     & mixed-AI & 27.60 & 4.00  & 56.31  & -  & 22.50 & 4.00  & 26.87  & - & 16.2 & 4 & 26.31 & - & 20.4 & 4 & 26.57 & -  \\ 
     &\cellcolor{gray!15}$\text{Relative}\Delta$ & \cellcolor{gray!15}\textcolor{red}{-33.90}  & \cellcolor{gray!15}\textcolor{red}{-22.22}  & \cellcolor{gray!15}25.92  & \cellcolor{gray!15}\textcolor{red}{-10.07}  & \cellcolor{gray!15}19.64  & \cellcolor{gray!15}\textcolor{red}{-22.22}  & \cellcolor{gray!15}\textcolor{red}{-93.61}  & \cellcolor{gray!15}\textcolor{red}{-32.06} & \cellcolor{gray!15}56.95  & \cellcolor{gray!15}0.00 & \cellcolor{gray!15}\textcolor{red}{-104.29} & \cellcolor{gray!15}\textcolor{red}{-15.78} & \cellcolor{gray!15}32.1  & \cellcolor{gray!15}0.00 & \cellcolor{gray!15}\textcolor{red}{-96.08} & \cellcolor{gray!15}\textcolor{red}{-21.33}  \\ 
     & \cellcolor{gray!15}$\text{Normalized}\Delta$ & \cellcolor{gray!15}\textcolor{red}{-34.89}  & \cellcolor{gray!15}\textcolor{red}{-22.22}  & \cellcolor{gray!15}13.06  & \cellcolor{gray!15}\textcolor{red}{-14.68}  & \cellcolor{gray!15}34.67  & \cellcolor{gray!15}\textcolor{red}{-22.22}  & \cellcolor{gray!15}\textcolor{red}{-71.32}  & \cellcolor{gray!15}\textcolor{red}{-19.62} & \cellcolor{gray!15}61.99  & \cellcolor{gray!15}0.00 & \cellcolor{gray!15}\textcolor{red}{-86.22} & \cellcolor{gray!15}\textcolor{red}{-8.08} & \cellcolor{gray!15}45.86  & \cellcolor{gray!15}0.00 & \cellcolor{gray!15}\textcolor{red}{-72.5} & \cellcolor{gray!15}\textcolor{red}{-8.88} \\  \bottomrule 
  \end{tabular}
  }
  \begin{minipage}{\textwidth}
  \footnotesize
  \raggedright
  \textit{\\ * The trends for R@5 and R@10 are similar to R@1.}
  \end{minipage}
\end{table*}

\section{Video Source Bias Assessment}
\label{sec:3}
In this section, we select three retrieval models and conduct text-to-video retrieval on the benchmark we construct, in order to evaluate the impact of incorporating AI-generated videos into the video library on retrieval performance. Moreover, through experiments involving AI-generated videos in both the test and training sets, we observe that the retrieval models tend to rank AI-generated videos higher in the retrieval results.

\subsection{Text-Video Retrieval Models}
To assess the source bias of retrieval models, it is first necessary to select appropriate models. We choose three distinct open-source video-text retrieval models with robust zero-shot retrieval capabilities to examine the source bias of AI-generated videos, including:

\textbf{(1) Frozen in Time~\cite{bain2021frozen}}: This model uses joint video and image encoders for end-to-end retrieval. 
It employs a Space-Time Transformer Encoder, which processes both image and video data flexibly, treating images as "frozen" snapshots of videos during training.

\textbf{(2) ALPRO~\cite{li2022align}}: By sparsely sampling video frames, ALPRO achieves effective cross-modal alignment without explicit object detectors. 
It introduces a Video-Text Contrastive Loss for aligning video and text features, simplifying cross-modal interaction modeling. 
Additionally, ALPRO features a Prompting Entity Modeling task for fine-grained alignment of visual regions and textual entities via self-supervision.

\textbf{(3) InternVideo~\cite{wang2022internvideo}}: This model combines generative and discriminative self-supervised learning strategies to optimize video representations. 
It uses Masked Video Modeling and Video-Language Contrastive Learning as pretraining tasks, with a learnable coordination mechanism to integrate both types of video representations.

All datasets and retrieval models demonstrated statistically significant results (p < 0.05). Detailed statistical analyses are provided in the Appendix~\ref{app:significance}.

\begin{figure*}[ht]
\centering
\begin{subfigure}{0.16\textwidth}
        \includegraphics[width=\linewidth]{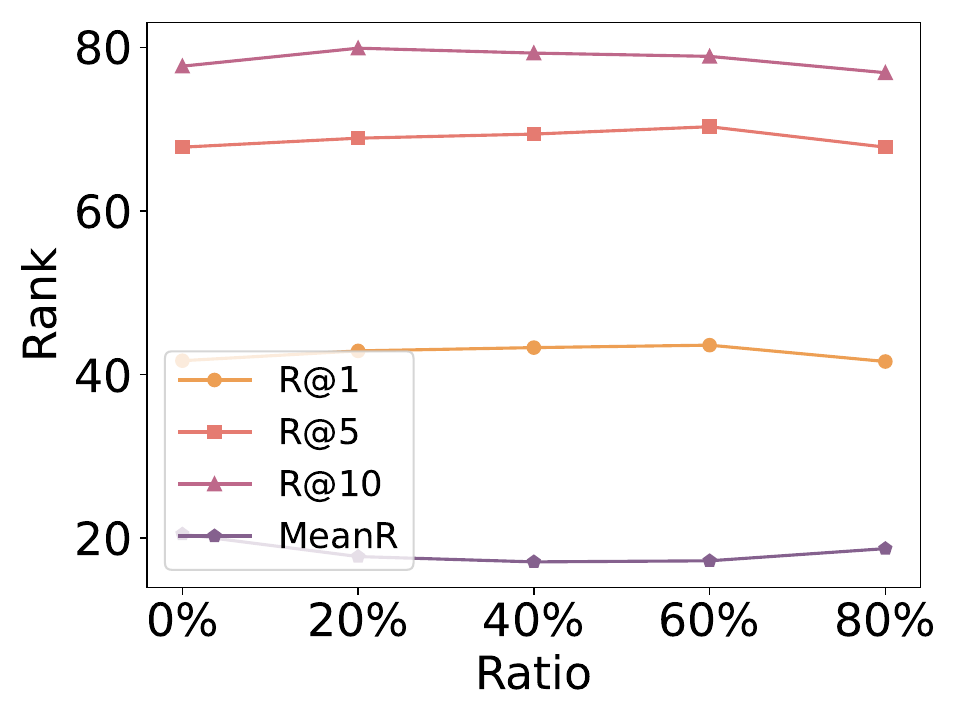}
        \caption*{\footnotesize(a) $\text{R@k}$ of Independent retrieval on real videos}
        \label{fig:REALR@k}
\end{subfigure}
\hfill
\begin{subfigure}{0.16\textwidth}
    \includegraphics[width=\linewidth]{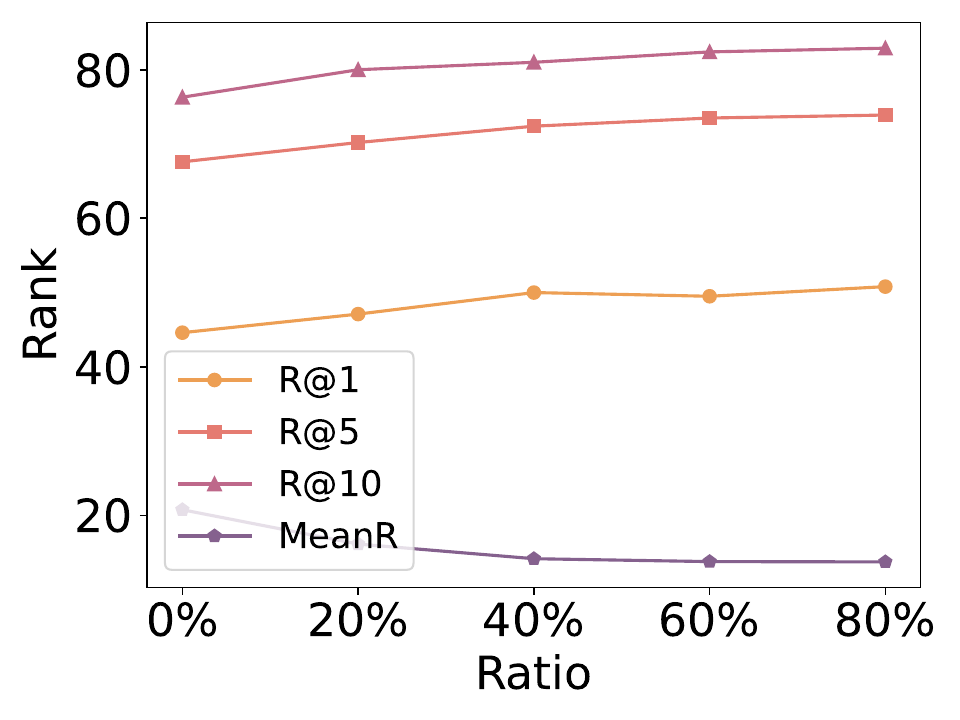}
    \caption*{\footnotesize(b) $\text{R@k}$ of Independent retrieval on AI videos}
    \label{fig:AIR@k}
\end{subfigure}
\hfill
\begin{subfigure}{0.16\textwidth}
    \includegraphics[width=\linewidth]{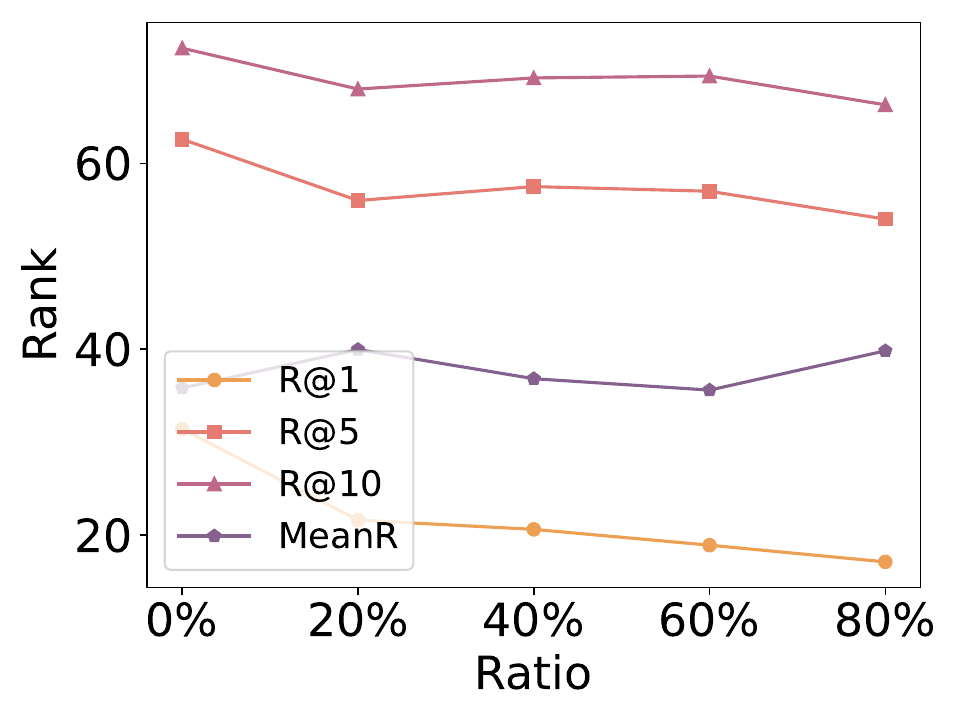}
    \caption*{\footnotesize(c) $\text{R@k}$ of real videos retrieval on mixed dataset}
    \label{fig:mixRealR@k}
    
\end{subfigure}
\hfill
\begin{subfigure}{0.16\textwidth}
    \includegraphics[width=\linewidth]{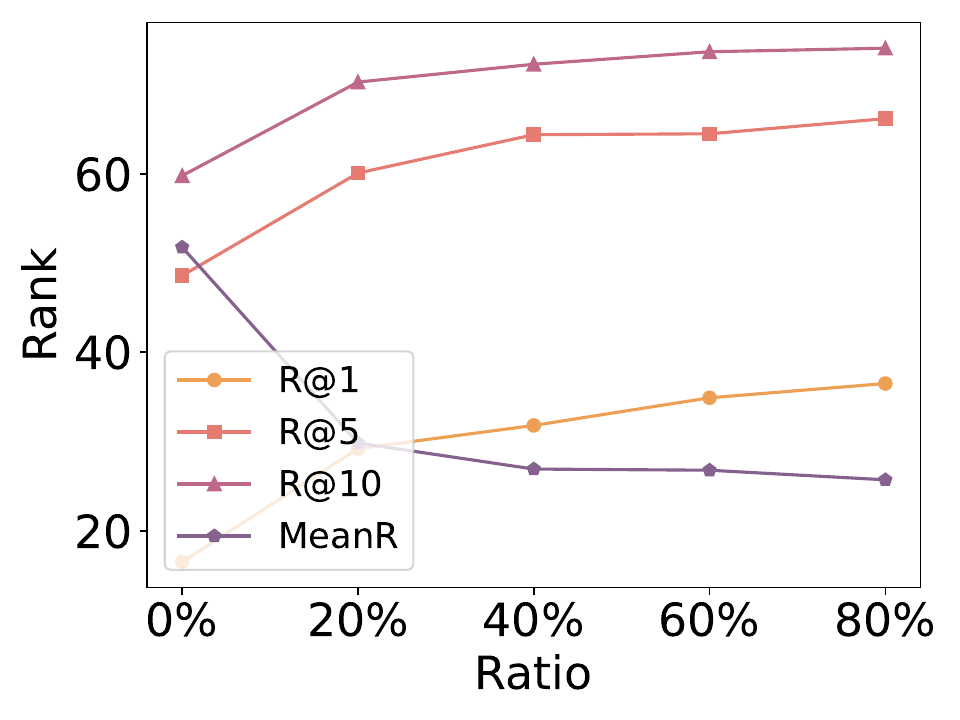}
    \caption*{\footnotesize(d) $\text{R@k}$ of AI videos retrieval on mixed dataset}
    \label{fig:mixAIR@k}
\end{subfigure}
\hfill
\begin{subfigure}{0.16\textwidth}
    \includegraphics[width=\linewidth]{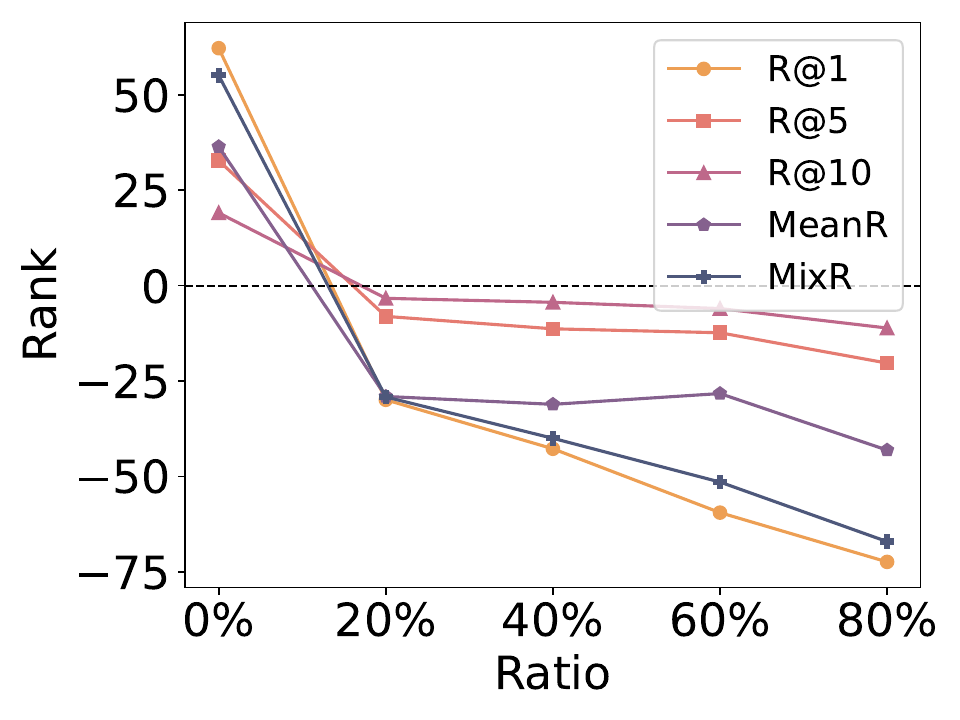}
    \caption*{\footnotesize(e) $\text{R@k}$ of $\text{Relative}\Delta$ on mixed dataset}
    \label{fig:relativedeltaR@k}
\end{subfigure}
\hfill
\begin{subfigure}{0.16\textwidth}
    \includegraphics[width=\linewidth]{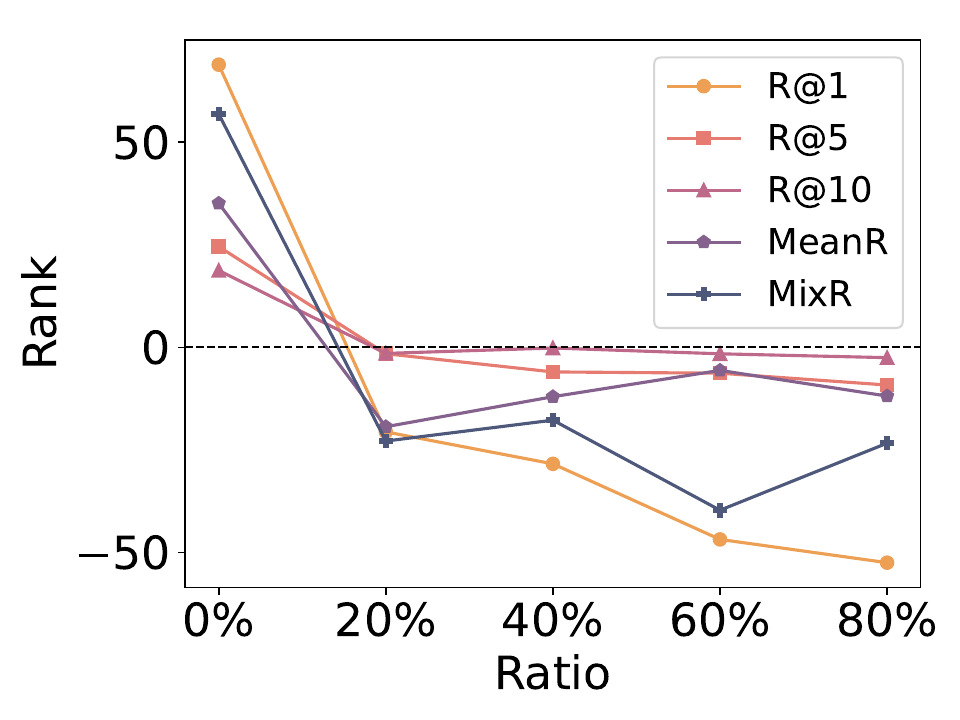}
    \caption*{\footnotesize(f) $\text{R@k}$ of $\text{Normalized}\Delta$ on mixed dataset}
    \label{fig:debiasdeltaR@k}
\end{subfigure}
\caption{Evaluation results on a training set containing a mix of AI-generated videos. We vary the proportion of AI-generated videos in the dataset (X-axis), while keeping the total number of training samples constant. The model is then tested on the OpenSora TextCond test set.}
\label{fig:mix-metric}
\end{figure*}

\subsection{Visual-Temporal Induced Source Bias}
\label{sec:3.2}
Experiments shown in Table~\ref{tab:main-exp-text} demonstrate that source bias is prevalent across various video retrieval models. Specifically, retrieval models tend to rank AI-generated videos higher than real videos, showing a clear preference for AI-generated content. Our specific key findings are:

(1) \textbf{Source bias is a widespread phenomenon, not specific to any particular video generation or retrieval model.} As shown in Table \ref{tab:main-exp-text}, for different video generation models, both $\text{Relative}\Delta$ and $\text{Normalized}\Delta$ values are generally negative. This suggests that vision-language models pre-trained on large video-text and image-text datasets tend to rank AI-generated videos higher.

(2) \textbf{Incorporating video or image segment information into the video generation process amplifies the Visual-Temporal Induced Source Bias.} As shown in Table~\ref{tab:main-exp-text}, we repeat the experiment on the OpenSora ImageCond and VideoExt datasets, where real video content is integrated into the generated videos. Our findings show that the retrieval metric gap narrows during individual retrieval, while the $\text{Relative}\Delta$ and $\text{Normalized}\Delta$ values increase.

(3) \textbf{When AI-generated videos are included in a video library, source bias significantly influences both users' initial impressions and their overall satisfaction with the search results ~\cite{al2010review}.} Specifically, the presence of source bias affects multiple retrieval metrics, including $\text{R@1}$, $\text{MeanR}$, and $\text{MedR}$. $\text{R@1}$ indicates the content users encounter first, $\text{MeanR}$ measures overall retrieval performance, and $\text{MedR}$ offers a more robust metric by minimizing the influence of outliers, though it is less sensitive to minor ranking changes. $\text{MixR}$ provides a comprehensive assessment of the impact of source bias on retrieval performance.

\subsection{More Serious Bias Caused by Training}
As AI-generated videos become increasingly pervasive on the Internet, they inevitably become integrated into the training datasets of video retrieval models. Our findings indicate that including AI-generated videos in the training set leads retrieval models to rank these videos higher, thereby exacerbating source bias.

To investigate how AI-generated videos affect model training, we constructed a mixed training set by replacing portions of real videos in the MSR-VTT dataset with their AI-generated counterparts. As shown in Figure~\ref{fig:mix-metric}, we fine-tuned the InternVideo model and observed that the model's preference for AI-generated videos increases as their proportion rises from 0\% to 20\%, 40\%, 60\%, and 80\%. Moreover, the Visual-Temporal Induced Source Bias becomes more pronounced with the growing presence of AI-generated content.

Taking the 20\% scenario as an example, the experimental results shown in Figure~\ref{fig:mix-metric} demonstrate that fine-tuning retrieval models on a real video training set improves retrieval performance compared to the original model, while significantly reducing source bias. Specifically, in terms of the $\text{Normalized}\Delta$ metrics, $\text{R@1}$ increases by 49.29 points, $\text{MeanR}$ by 106.45 points, and $\text{MixR}$ by 76.53 points. These improvements suggest that incorporating real data into the fine-tuning set can effectively mitigate source bias.

Moreover, fine-tuning with a mixed training set containing 20\% AI-generated videos also enhances the model’s retrieval performance. However, when compared to fine-tuning using only real videos, the source bias increases substantially. The $\text{Normalized}\Delta$ metrics indicate a decrease of 89.52 in $\text{R@1}$, 54.47 in $\text{MeanR}$, and 79.74 in $\text{MixR}$. These findings imply that, while the overall retrieval performance may not differ dramatically, the model becomes more inclined to retrieve AI-generated videos as their proportion in the training set increases. Even a 20\% inclusion of AI-generated videos significantly impacts source bias.

\label{sec:3.3}

\section{Causes of Video Source Bias}

\label{sec:4}
% In this section, we explore the causes of source bias in AI-generated videos. We identify that source bias stems from two key sources of information embedded in these videos: visual and temporal data. We investigate the origins of these biases using methods such as frame shuffling and single-frame retrieval.

In this section, we analyze the origins of source bias in AI-generated videos by considering two key components: visual and temporal information. We use frame shuffling and single-frame retrieval to isolate and assess the influence of each component.

\subsection{Bias Induced by Visual Information}
\label{sec:4.1}
Our experiments show that the visual content of a video plays a major role in source bias. We also find that real videos contain more diverse and richer temporal information. As shown in Table \ref{tab:flow_entropy}. To study this, we kept the visual content unchanged and manipulated the temporal sequence by rearranging the frames in random or reverse order, changing the temporal flow without affecting visual quality.

\begin{table}[h]
\centering
\caption{Comparative optical flow entropy in OpenSora TextCond: Real vs. AI-generated videos.}
\resizebox{0.4\textwidth}{!}{
\begin{tabular}{lcc}
\toprule
 & Real Videos & AI-Generated Videos \\ 
\midrule
Mean Entropy & 2.9280 & 1.7453 \\
Higher Entropy Count & 851 & 149 \\ 
\bottomrule
% \hline
\end{tabular}}
\begin{minipage}{\textwidth}
\footnotesize
\raggedright
\textit{\\ * Higher entropy indicates richer temporal information.}
\end{minipage}
\label{tab:flow_entropy}
\end{table}

We conducted experiments on OpenSora TextCond and real videos, keeping the original frame rate. The results (see Table~\ref{tab:main-exp-random}) show that even when the temporal order is altered, visual features still strongly influence retrieval bias. This supports the idea that visual information in AI-generated videos makes them more likely to be favored in retrieval tasks.

Furthermore, we investigate two scenarios to assess the impact of temporal information on retrieval bias: first, shuffling only the frames of AI-generated videos (Random-only-AI), and second, shuffling the frames of both real and AI-generated videos (Random). The first scenario tests if temporal information in AI-generated videos contributes to source bias while keeping the visual content of real videos intact. The second scenario compares how temporal sequence affects retrieval performance for both video types. Our findings show that shuffling only the frames of AI-generated videos reduces retrieval accuracy. However, when frames from both real and AI-generated videos are shuffled, the $\text{Normalized}\Delta$ significantly decreases.

\begin{table}
   \caption{The retrieval performance of Text-Video Retrieval Models on the OpenSora TextCond dataset is assessed following the shuffling of video frame order. }
   % 'Random' denotes the shuffling of both real and AI-generated videos, while 'Random-only-AI' refers to shuffling only the AI-generated videos. For the definitions of $\text{Relative}\Delta$ and $\text{Normalized}\Delta$, please refer to Table~\ref{tab:main-exp-text}.
  \label{tab:main-exp-random}
  \centering
  \large
  \setlength{\tabcolsep}{2pt}
  \resizebox{0.45\textwidth}{!}{%
  \begin{tabular}{cccccccccc}
  \toprule
    \multicolumn{2}{c}{\makecell{\textbf{Dataset}}}  & 
    \multicolumn{4}{c}{\makecell{\textbf{Random}}} &
    \multicolumn{4}{c}{\makecell{\textbf{Random-only-AI-videos}}} 
    \\
     Model & \text{Metric} & \text{R@1} & \text{MedR} & \text{MeanR} & \text{MixR} & \text{R@1} & \text{MedR} & \text{MeanR} & \text{MixR} \\
     \midrule
     \multirow{6}{*}{\makecell[c]{Alpro}} & REAL & 23.5 & 9 & 53.71 & - & 24.1 & 8 & 49.61 & - \\
      & AI & 37.0 & 3 & 28.56 & - & 37.0 & 3 & 28.56 & -  \\
     & mixed-REAL & 9.7 & 16 & 90.49 & - & 10.3  & 13 & 84.72 & - \\
     & mixed-AI & 25.8  & 6 & 71.15 & - & 25.9 & 6 & 69.77 & - \\
     & \cellcolor{gray!15}$\text{Relative}\Delta$ & \cellcolor{gray!15}\textcolor{red}{-90.7} & \cellcolor{gray!15}\textcolor{red}{-90.91} & \cellcolor{gray!15}\textcolor{red}{-23.93} & \cellcolor{gray!15}\textcolor{red}{-68.51} & \cellcolor{gray!15}\textcolor{red}{-86.19} & \cellcolor{gray!15}\textcolor{red}{-73.68} & \cellcolor{gray!15}\textcolor{red}{-19.35} & \cellcolor{gray!15}\textcolor{red}{-59.74} \\
     & \cellcolor{gray!15}$\text{Normalized}\Delta$ & \cellcolor{gray!15}\textcolor{red}{-46.07} & \cellcolor{gray!15}13.44 & \cellcolor{gray!15}37.58 & \cellcolor{gray!15}1.65 & \cellcolor{gray!15}\textcolor{red}{-43.96} & \cellcolor{gray!15}21.56 & \cellcolor{gray!15}34.85 & \cellcolor{gray!15}4.15 \\ 
     \multirow{6}{*}{\makecell[c]{Frozen}} & REAL & 20.5 & 9 & 56.19 & - & 22.9  & 8 & 49.81 & -  \\
     & AI & 30.7 & 4 & 32.57 & - & 30.7 & 4 & 32.57 & - \\
     & mixed-REAL & 7.2 & 19 & 102.92 & - & 9.3 & 16 & 90.58 & - \\
     & mixed-AI & 24.3 & 7 & 74.01 & - & 24 & 8 & 74.48 & -  \\
     & \cellcolor{gray!15}$\text{Relative}\Delta$ & \cellcolor{gray!15}\textcolor{red}{-108.57}  & \cellcolor{gray!15}\textcolor{red}{-92.31} & \cellcolor{gray!15}\textcolor{red}{-32.69} & \cellcolor{gray!15}\textcolor{red}{-77.86} & \cellcolor{gray!15}\textcolor{red}{-88.29} & \cellcolor{gray!15}\textcolor{red}{-66.67} & \cellcolor{gray!15}\textcolor{red}{-19.51} & \cellcolor{gray!15}\textcolor{red}{-58.16}  \\ 
    & \cellcolor{gray!15}$\text{Normalized}\Delta$ & \cellcolor{gray!15}\textcolor{red}{-68.73} & \cellcolor{gray!15}\textcolor{red}{-12.31} & \cellcolor{gray!15}20.83 & \cellcolor{gray!15}\textcolor{red}{-20.07} & \cellcolor{gray!15}\textcolor{red}{-59.19}  & \cellcolor{gray!15}2.9 & \cellcolor{gray!15}22.6 & \cellcolor{gray!15}\textcolor{red}{-11.23}  \\ 
    \multirow{6}{*}{\makecell[c]{Intern\\Video}} & REAL & 39.6 & 2 & 24.56 & - & 40.6 & 2 & 22.27 & -  \\
     & AI & 47 & 2 & 18.04 & - & 47 & 2 & 18.04 & -  \\
     & mixed-REAL & 22.9 & 6 & 86.24 & - & 28.3 & 4 & 75.43 & -  \\
     & mixed-AI &
    28 & 3 & 25.37 & - & 20.8 & 4 & 27.36 & -  \\
     & \cellcolor{gray!15}$\text{Relative}\Delta$ &
    \cellcolor{gray!15}\textcolor{red}{-20.04} & \cellcolor{gray!15}\textcolor{red}{-66.67} & \cellcolor{gray!15}\textcolor{red}{-109.07} & \cellcolor{gray!15}\textcolor{red}{-65.26} & \cellcolor{gray!15}30.55 & \cellcolor{gray!15}0 & \cellcolor{gray!15}\textcolor{red}{-93.53} & \cellcolor{gray!15}\textcolor{red}{-20.99}  \\
     &  \cellcolor{gray!15}$\text{Normalized}\Delta$ & \cellcolor{gray!15}\textcolor{red}{-2.95} & \cellcolor{gray!15}\textcolor{red}{-66.67} & \cellcolor{gray!15}\textcolor{red}{-78.14} & \cellcolor{gray!15}\textcolor{red}{-49.25} & \cellcolor{gray!15}45.16 & \cellcolor{gray!15}0 & \cellcolor{gray!15}\textcolor{red}{-72.32} & \cellcolor{gray!15}\textcolor{red}{-9.05}  \\ \bottomrule 
  \end{tabular}
  }
\end{table}

\begin{table}
  \belowrulesep=0pt
  \aboverulesep=0pt
  \caption{Model performance on single-frame retrieval using the OpenSora TextCond dataset.}
  \label{tab:singleframe}
  \centering
  \setlength{\tabcolsep}{0.8mm}
  \renewcommand{\arraystretch}{0.9}
  \resizebox{0.40\textwidth}{!}{
  \begin{tabular}{llcccccc}
    \toprule
     Model & \text{Metric} & \text{R@1} & \text{R@5} & \text{R@10} & \text{MedR} & \text{MeanR} & \text{MixR} \\
    \midrule
     \multirow{6}{*}{\makecell[c]{Alpro}} & REAL & 32.2 & 52.2 & 61.5 & 5 & 49.74 & -\\
     & AI & 33.3 & 56.9 & 66.6 & 3 & 37.69 & -\\
     & mixed-REAL & 16.3 & 42.5 & 52.4 & 9 & 83.66 & -\\ 
     & mixed-AI & 22.4 & 47.4 & 57.1 & 6 & 96.41 & -\\ 
     & \cellcolor{gray!15}$\text{Relative}\Delta$ & \cellcolor{gray!15}\textcolor{red}{-31.52} & \cellcolor{gray!15}\textcolor{red}{-10.9} & \cellcolor{gray!15}\textcolor{red}{-8.58} & \cellcolor{gray!15}\textcolor{red}{-40} & \cellcolor{gray!15}14.16 & \cellcolor{gray!15}\textcolor{red}{-19.12} \\
     & \cellcolor{gray!15}$\text{Normalized}\Delta$ & \cellcolor{gray!15}\textcolor{red}{-28.16} & \cellcolor{gray!15}\textcolor{red}{-3.35} & \cellcolor{gray!15}0.04 & \cellcolor{gray!15}13.33 & \cellcolor{gray!15}41.89 & \cellcolor{gray!15}9.02  \\ 
    \multirow{6}{*}{\makecell[c]{Frozen}} & REAL & 34.7 & 56.4 & 65.4 & 4 & 42.16 & - \\
     & AI & 37.7 & 61.6 & 69.3 & 3 & 30.00 & - \\
     & mixed-REAL &
    16.8 & 45.6 & 55.4 & 7 & 70.95 & - \\
     & mixed-AI &
    26.3 & 51.6 & 60.6 & 5 & 77.42 & - \\
     & \cellcolor{gray!15}$\text{Relative}\Delta$ & \cellcolor{gray!15}\textcolor{red}{-44.08} & \cellcolor{gray!15}\textcolor{red}{-12.35} & \cellcolor{gray!15}\textcolor{red}{-8.97} & \cellcolor{gray!15}\textcolor{red}{-33.33} & \cellcolor{gray!15}8.72 & \cellcolor{gray!15}\textcolor{red}{-22.9}  \\ 
    & \cellcolor{gray!15}$\text{Normalized}\Delta$ & \cellcolor{gray!15}\textcolor{red}{-35.79} & \cellcolor{gray!15}\textcolor{red}{-3.91} & \cellcolor{gray!15}\textcolor{red}{-0.16} & \cellcolor{gray!15}\textcolor{red}{-2.56} & \cellcolor{gray!15}42.66 & \cellcolor{gray!15}1.44 \\ 
    \multirow{6}{*}{\makecell[c]{Intern\\Video}} & REAL & 32.2 & 52.2 & 61.5 & 5 & 49.74  & - \\
     & AI & 41.6 & 66 & 75.3 & 2 & 23.33 & - \\
     & mixed-REAL &
    15.5 & 44 & 53.2 & 9 & 83.14 & - \\
     & mixed-AI &
    28.1 & 54.3 & 63.8 & 4 & 60.12  & - \\
     & \cellcolor{gray!15}$\text{Relative}\Delta$ & \cellcolor{gray!15}\textcolor{red}{-57.8} & \cellcolor{gray!15}\textcolor{red}{-20.96} & \cellcolor{gray!15}\textcolor{red}{-18.12} & \cellcolor{gray!15}\textcolor{red}{-76.92} & \cellcolor{gray!15}\textcolor{red}{-32.13} & \cellcolor{gray!15}\textcolor{red}{-55.62}  \\
     & \cellcolor{gray!15}$\text{Normalized}\Delta$ & \cellcolor{gray!15}\textcolor{red}{-32.33} & \cellcolor{gray!15}3.13 & \cellcolor{gray!15}5.23 & \cellcolor{gray!15}15.39 & \cellcolor{gray!15}40.66 & \cellcolor{gray!15}7.91   \\ 
    \bottomrule 
  \end{tabular}
  }
\end{table}

\subsection{Bias Induced by Temporal Information}

\label{sec:4.2}

In this section, we find that temporal information causes AI-generated videos to rank higher in the retrieval list, not just in the first position. When retrieval is based on a single frame, the video is treated as a static image, and the system relies only on the visual content of that frame, ignoring temporal sequence. This setup helps us analyze how the visual content influences retrieval results and potential biases when temporal and multi-frame information is absent.

The experimental results (Table~\ref{tab:singleframe}) show that Visual-Temporal Induced Source Bias still exists, but is mainly observed in the $\text{R@1}$ metric. In the retrieval results across the three models, source bias is prominent in $\text{R@1}$ but nearly absent in $\text{MeanR}$ and $\text{MedR}$. This suggests that, when using a single frame, the model’s bias is strongest in its preference for the first retrieved frame.

\section{Mitigating and Visualizing Bias}
\label{sec:5}
In this section, we propose a method using contrastive learning to mitigate this bias. By incorporating AI-generated videos into the contrastive learning training set, we fine-tune the model to increase the likelihood of retrieving real videos while reducing the likelihood of retrieving AI-generated ones. Additionally, we extract a debiasing vector from the model, which can be applied to other video encoding vectors to further reduce the Visual-Temporal Induced Source Bias in the retrieval system. This vector can also be used to visualize the bias.

\subsection{Debiased Model Training}
\label{sec:5.1}

We use the OpenSora TextCond (Train) dataset from Section~\ref{sec:2} to train the debiasing model with a contrastive learning approach.

\textbf{Notation:} A real video corresponds to both an AI-generated video and a caption. 
For a video-text pair, if the video is AI-generated, we represent it as $(V_{\scriptscriptstyle \text{G}}, C)$, and if it is a real video, we represent it as $(V_{\scriptscriptstyle \text{R}}, C)$. 
In retrieval, the model first samples $f$ frames from the video, represented as $I_j$ ($j \in [1, f]$). 
It then uses the pre-trained image encoder $E_{\scriptscriptstyle \text{I}}$ and text encoder $E_{\scriptscriptstyle \text{C}}$ to encode the images and text into vectors $h_{I_j}$ and $h_{\scriptscriptstyle \text{C}}$. 
These vectors are subsequently input into the video retrieval model $E_{\scriptscriptstyle \text{V}}$, which computes the final video-text similarity $r_{\scriptscriptstyle \text{VC}}$, expressed as:
\begin{align}
    r_{\scriptscriptstyle \text{VC}} =
    E_{\scriptscriptstyle \text{V}}([h_{I_1}...h_{I_f}],
    h_{\scriptscriptstyle \text{C}},
    \theta_{\scriptscriptstyle \text{VC}}),  \\
    h_{I_j} = E_{\scriptscriptstyle \text{I}}(I_j,\theta_{\scriptscriptstyle \text{I}}),\ \ h_{\scriptscriptstyle \text{C}} = E_{\scriptscriptstyle \text{C}}(C,\theta_{\scriptscriptstyle \text{C}}).
    \label{eq:rh}
\end{align}

\textbf{Loss Construction:} The optimization objective is as follows. Let $y$ represent the label: when $y = 1$, the video corresponds to the text, and when $y = 0$, they do not correspond. $\mathcal{L}$ denotes the loss function, which minimizes the distance between the image and its corresponding text embedding vectors:
\begin{equation}
\theta_{\scriptscriptstyle \text{VC}} = \arg\min_{\theta_{\text{vc}}}\ \mathcal{L}(r_{\scriptscriptstyle \text{VC}},y,\theta_{\scriptscriptstyle \text{VC}}).
\label{eq:thetaVC}
\end{equation}

During the training of the debiased model, we aim to enable the retrieval model to more easily retrieve the corresponding real videos based on the text, while avoiding AI-generated videos. For each text, we have a real video-AI-generated video-text triplet $(V_{\scriptscriptstyle \text{R}}, V_{\scriptscriptstyle \text{G}}, C_i)$. After image sampling, we use the image encoder and text encoder to encode them into vectors $h_{Rf}$, $h_{Gf}$, and $h_{Ci}$, respectively. A contrastive loss function is introduced in the debiased model:
\begin{equation}
\Delta r = E_{\scriptscriptstyle \text{V}}([h_{G_1} \dots h_{G_f}],h,\theta_{\scriptscriptstyle \text{VC}}) \\ -E_{\scriptscriptstyle \text{V}}([h_{R_1} \dots h_{R_f}],h,\theta_{\scriptscriptstyle \text{VC}}),
\end{equation}
% \Delta r = _{(G,R,C)}
where $\Delta r$ measures the difference in scores assigned by the model between two videos, providing a comprehensive way to assess the consistent invisible bias of the model toward real and AI-generated videos. This helps guide the model in reducing bias against generated videos. Additionally, when $\Delta r < 0$, we do not apply this loss function, ensuring that during contrastive learning, the model still favors generated videos while increasing the likelihood of retrieving real videos. The overall training objective is:
\begin{align}
     \theta_{\scriptscriptstyle \text{VC}} = \arg\min_{\theta_{\text{vc}}}\ \mathcal{L}(r_{\scriptscriptstyle  \text{VC}},y,\theta_{\scriptscriptstyle \text{VC}}) + \Delta r.
     \label{thetaVC2}
\end{align}

The model training results, as shown in Table~\ref{tab:debias}, indicate that the debiased model is more likely to rank real videos at the top of the list, significantly reducing the model’s source bias.

\begin{table}
  \belowrulesep=0pt
  \aboverulesep=0pt
  \caption{The performance of the InternVideo model after debiasing fine-tuning using contrastive learning. Fine-tuning is performed on the OpenSora TextCond dataset, and testing is conducted on three different datasets.}
  \label{tab:debias}
  \centering
  \setlength{\tabcolsep}{0.8mm}
  \renewcommand{\arraystretch}{0.9}
  \small
  \resizebox{0.40\textwidth}{!}{
  \begin{tabular}{llcccccc}
  \toprule
    \multicolumn{8}{c}{\cellcolor{white}\makecell[l]{\hspace{5em}\textbf{InternVideo Contrastive-Debias}}}
    \\
    % \midrule
     Dataset & \text{Metric} & \text{R@1} & \text{R@5} & \text{R@10} & \text{MedR} & \text{MeanR} & \text{MixR} \\
    \midrule
    \multirow{6}{*}{\makecell[c]{OpenSora\\TextCond}} & REAL & 41.2 & 66.5 & 76 & 2 & 20.436 & - \\
     & AI & 41.5 & 64.5 & 73.9 & 2 & 23.662 & - \\
     & mixed-REAL & 41.2 & 66.5 & 76 & 2 & 23.534 & - \\ 
     & mixed-AI & 0 & 0.2 & 0.5 & 224 & 293.688 & - \\ 
     & \cellcolor{gray!15}$\text{Relative}\Delta$ & \cellcolor{gray!15}200 & \cellcolor{gray!15}198.8 & \cellcolor{gray!15}197.39 & \cellcolor{gray!15}196.46 & \cellcolor{gray!15}170.32 & \cellcolor{gray!15}188.93 \\
     & \cellcolor{gray!15}$\text{Normalized}\Delta$ & \cellcolor{gray!15}200.73 & \cellcolor{gray!15}195.28 & \cellcolor{gray!15}194.34 & \cellcolor{gray!15}196.46 & \cellcolor{gray!15}155.52 & \cellcolor{gray!15}184.24 \\
    \multirow{6}{*}{\makecell[c]{OpenSora\\ImageCond}} & REAL & 41.2 & 66.5 & 76 & 2 & 20.436 & - \\
     & AI & 35.3 & 63.1 & 74.2 & 3 & 22 & - \\
     & mixed-REAL & 41.2 & 66.5 & 76 & 2 & 23.727 & - \\ 
     & mixed-AI & 0 & 0.3 & 0.4 & 247 & 304.562 & - \\ 
     & \cellcolor{gray!15}$\text{Relative}\Delta$ & \cellcolor{gray!15}200 & \cellcolor{gray!15}198.2 & \cellcolor{gray!15}197.91 & \cellcolor{gray!15}196.79 & \cellcolor{gray!15}171.09 & \cellcolor{gray!15}189.29 \\
     & \cellcolor{gray!15}$\text{Normalized}\Delta$ & \cellcolor{gray!15}184.58 & \cellcolor{gray!15}189.5 & \cellcolor{gray!15}192.66 & \cellcolor{gray!15}152.35 & \cellcolor{gray!15}163.63 & \cellcolor{gray!15}166.85 \\
    \multirow{6}{*}{\makecell[c]{CogVideoX\\TextCond}} & REAL & 41.2 & 66.5 & 76 & 2 & 20.436 & - \\
     & AI & 37.1 & 61.4 & 70.9 & 3 & 24.279 & - \\
     & mixed-REAL & 41.2 & 66.5 & 76 & 2 & 24.13 & - \\ 
     & mixed-AI & 0 & 2.9 & 7.5 & 99.5 & 183.991 & - \\ 
     & \cellcolor{gray!15}$\text{Relative}\Delta$ & \cellcolor{gray!15}200 & \cellcolor{gray!15}183.29 & \cellcolor{gray!15}164.07 & \cellcolor{gray!15}192.12 & \cellcolor{gray!15}153.62 & \cellcolor{gray!15}181.91 \\
     & \cellcolor{gray!15}$\text{Normalized}\Delta$ & \cellcolor{gray!15}189.53 & \cellcolor{gray!15}173.32 & \cellcolor{gray!15}156.1 & \cellcolor{gray!15}147.68 & \cellcolor{gray!15}136.24 & \cellcolor{gray!15}157.82\\ 
    \bottomrule 
  \end{tabular}
  }
\end{table}

\begin{figure}
\centering
\includegraphics[width=0.5\linewidth]{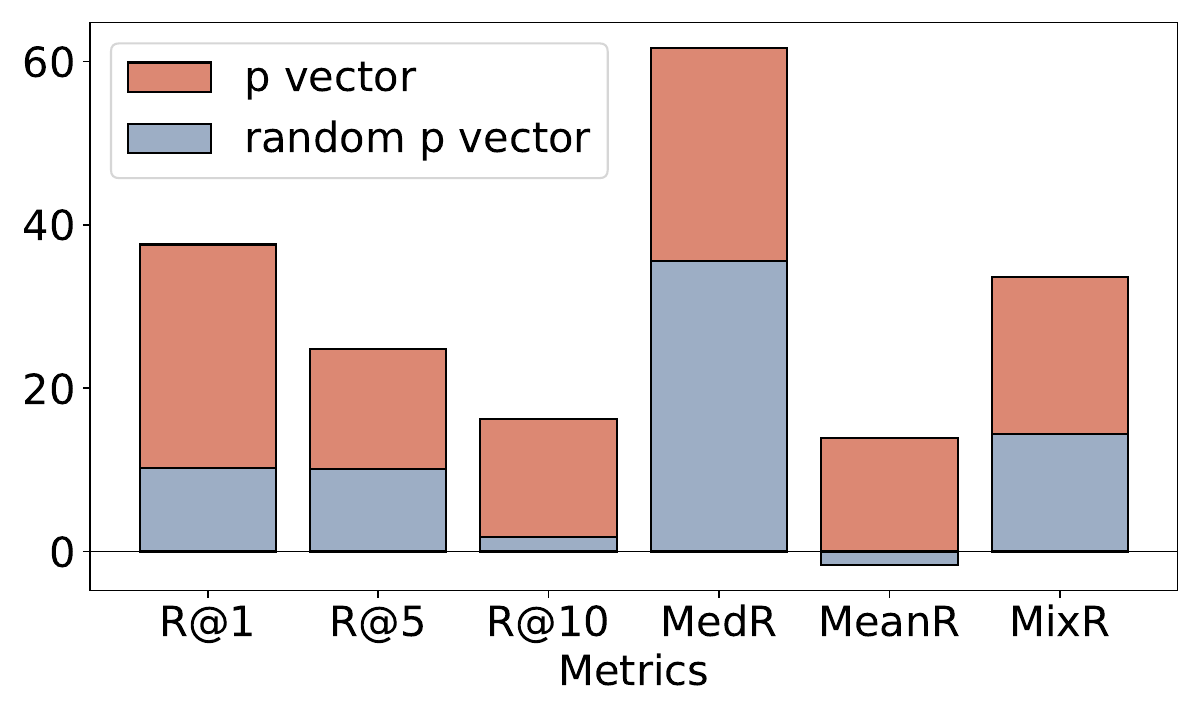}
\caption{Changes in retrieval metrics of the model ($\Delta$) after altering the vector representations of videos using p-vectors and random p-vectors.}
\label{fig:p}
\end{figure}

\subsection{Visual and Temporal Vectors Visualization}
\label{sec:5.2}
In this section, we use the debiased model from the previous section to analyze the source of bias and visualize the invisible source bias. Given that the retrieval model exhibits a general preference for AI-generated videos, which is reversed by the debiased model, we analyze the video embeddings after debiasing to explore this reversal.

\textbf{Notation:} After obtaining the debiased model, the debiased video encoder is denoted as $E_v^d$. We can obtain the original video embedding $h_v = [h_1, h_2, \dots, h_n]$, the debiased video embedding $h_v^d = [h_1^d, h_2^d, \dots, h_n^d]$, and the text embedding $C = [c_1, c_2, \dots, c_n]$.

\textbf{Visualization Bias:} We define a vector $p$ to represent the difference between the debiased video embedding and the original video embedding, capturing the shift in the video representation after debiasing:
\begin{align}
     p =  [p_1,p_2,...,p_n]
       =  [h_{1}^{d}-h_1,h_{2}^{d}-h_2,...,h_{n}^{d}-h_n].
\label{eq:p}
\end{align}

\begin{figure}
\centering
\begin{subfigure}{0.2\textwidth}
        \includegraphics[width=\linewidth]{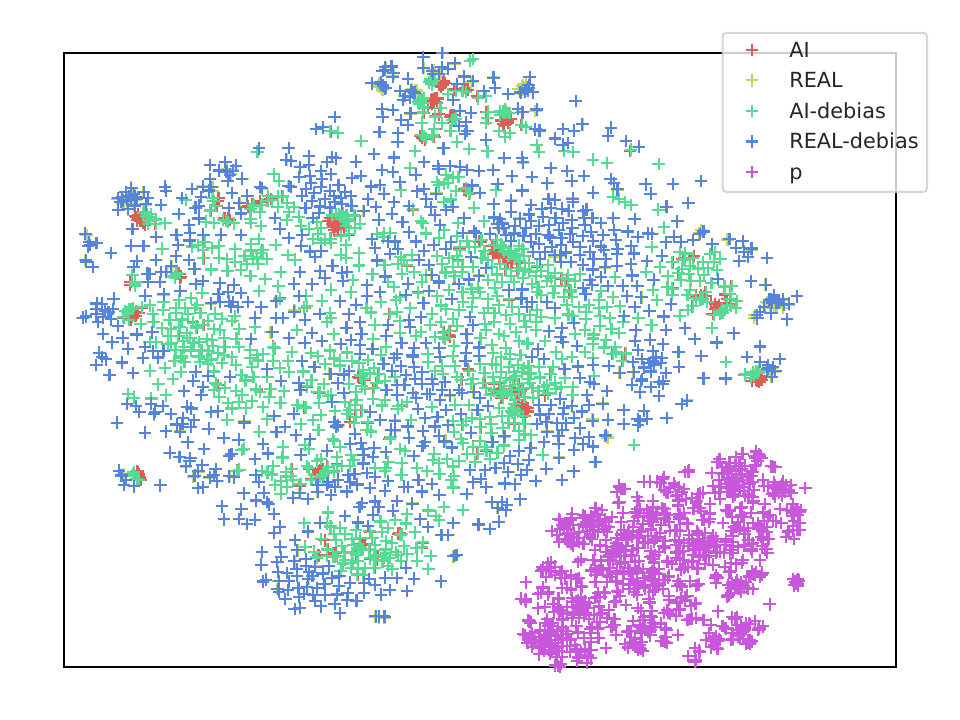}
        \caption*{(a) Original Temporal Information t-SNE Visualization}
        \label{fig:TSNE1}
\end{subfigure}
\hfill
\begin{subfigure}{0.2\textwidth}
    \includegraphics[width=\linewidth]{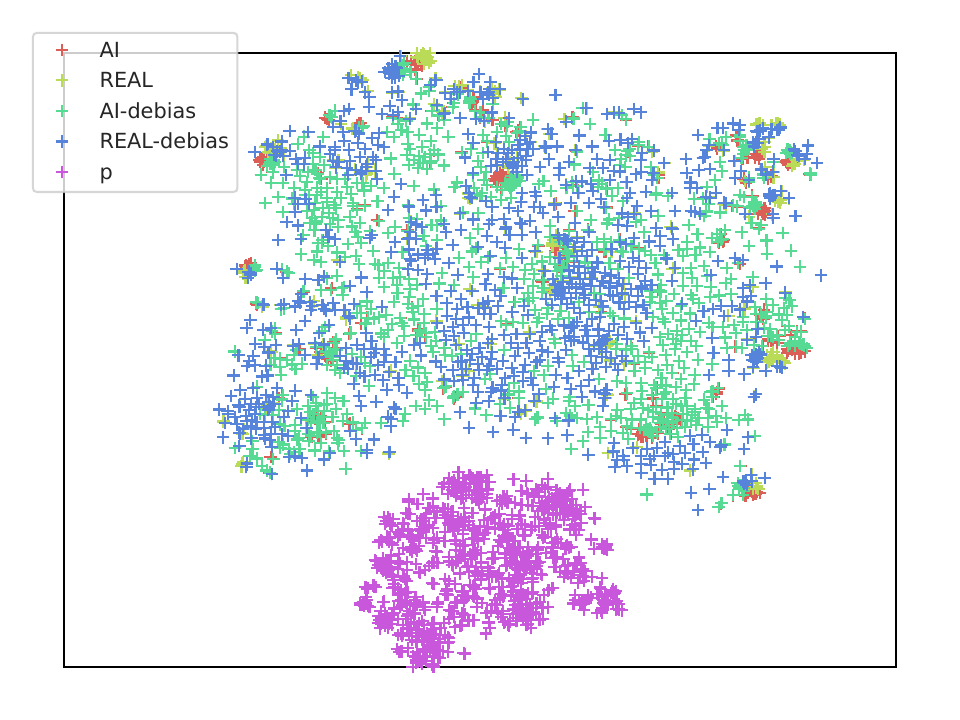}
    \caption*{(b) Scramble Temporal Information t-SNE Visualization}
    \label{fig:TSNE2}
\end{subfigure}
\hfill
\caption{t-SNE visualization of image representations and
transformations vector.}
\label{fig:TSNE}
\end{figure}

After performing t-SNE visualization on the vectors $p$, $h_v$, and $h_v^d$, as shown in Figure~\ref{fig:TSNE}, we observe that the vector $p$ forms a distinct clustering pattern. 
This indicates that the AI video generation model embeds additional, consistent information across generated videos. 
We identify these extra details as the direct cause of the Visual-Temporal Induced Source Bias. 
Three key observations emerged: 
(1) These extra details are generalizable, and incorporating them into real videos can increase their retrieval probability and reduce bias.
(2) Some of this additional information encodes temporal aspects, with certain details integrated into the generated videos through temporal sequences. 
(3) The extra information exhibits a high degree of consistency, as all AI-generated videos share a concentrated embedding of this additional information.
\begin{table}
\belowrulesep=0pt
  \aboverulesep=0pt
    \caption{The effect of adding the extracted $p$ or $p_{\text{random}}$ vector to the original video representation on retrieval performance:
    $\Delta > 0$ indicates that adding the vector (either $p$ or $p_{\text{random}}$) makes the retrieval model more likely to rank real videos higher,
    while $\Delta < 0$ suggests that the model tends to prioritize AI-generated videos.}
  \label{tab:p-debias}
  \centering
  \setlength{\tabcolsep}{0.8mm}
  \renewcommand{\arraystretch}{0.9}
  \small
  % \rowcolors{2}{white}{gray!15}
  \resizebox{0.4\textwidth}{!}{%
  \begin{tabular}{lcccccc}
  \toprule
    \multicolumn{7}{c}{\cellcolor{white}\makecell[l]{\hspace{5em}\textbf{InternVideo p-Debias}}}
    \\\midrule
     \text{Metric} & \text{R@1} & \text{R@5} & \text{R@10} & \text{MedR} & \text{MeanR} & \text{MixR} \\
    \midrule
    $\text{REAL}$ & -20.48 & -17.32 & -15.04 & -22.23 & -39.9 & -23.49 \\
    $\text{REAL}_{\text{p-debias}}$ & 17.12 & 7.54 & 1.21 & 39.4 & -26.0 & 10.17 \\
    $\Delta$ & 37.60 & 24.86 & 16.25 & 61.63 & 13.90 & 33.66 \\ \midrule
    \multicolumn{7}{c}{\cellcolor{white}\makecell[l]{\hspace{5em}\textbf{InternVideo p-random-Debias}}}
    \\
    \midrule
    \text{Metric} & \text{R@1} & \text{R@5} & \text{R@10} & \text{MedR} & \text{MeanR} & \text{MixR} \\
    \midrule
    $\text{REAL}$ & -51.36 & -12.0 & -7.06 & -35.56 & 0.42 & -28.83 \\
    $\text{REAL}_{\text{p-debias}}$ & -41.14 & -1.9 & -5.29 & -4.19 & 2.04 & -14.43 \\
    $\Delta$ & 10.22 & 10.1 & 1.77 &  35.56 & -1.62 & 14.40 \\
    \bottomrule 
  \end{tabular}
  }
\end{table}

For the first point, we design an experiment to investigate the effect of the debiasing model. Since the model adds a vector $p_i = h_{R}^d - h_R$ to each generated video $V_{G_i}$, we simplify the analysis by using the average of all $p_i$ vectors, denoted as $p_{avg}$, to represent the additional information introduced. Using the original video encoder, we obtain the raw representation of the real video $h_{R} = [h_{R_1}, h_{R_2}, \dots, h_{R_n}]$, and then add the computed $p_{avg}$ to each video representation, resulting in the biased real video representation $h_{R}^{pd} = [h_{R_1} + p_{avg}, h_{R_2} + p_{avg}, \dots, h_{R_n} + p_{avg}]$. The experimental results, as shown in Table~\ref{tab:p-debias}, demonstrate that inputting the new real video representation into the retrieval model leads to a reduction in bias across all evaluation metrics, particularly in the R@1 metric, where the bias is even reversed. This finding suggests that the additional information is universal; it can be applied not only to AI-generated videos but also directly to real videos, thereby enhancing their retrieval performance. This confirms that the vector $p$ plays a crucial role in introducing Visual-Temporal Induced Source Bias.

For the second point, we investigate whether the additional information includes temporal data. We input the Random OpenSora TextCond dataset into the model, generating and extracting the $p_{\text{random}}$ vector, which contains scrambled temporal information. This $p_{\text{random}}$ vector preserves the original visual content while introducing disrupted temporal sequences. When this vector is added to the real video representation, the experimental results, as shown in Table~\ref{tab:p-debias}, indicate a reduction in bias across the evaluation metrics. However, the effect is weaker compared to when real temporal information is included. Figure~\ref{fig:p} suggests that temporal information in AI-generated videos plays a significant role in Visual-Temporal Induced Source Bias, with additional information encoded within the temporal sequences.

For the third point, we visualize the vectors $p$, $h_G$, $h_R$, $h_{G}^d$, and $h_{R}^d$, along with the corresponding vectors from the random dataset, in two dimensions. The results are presented in Figure~\ref{fig:TSNE}. From the t-SNE plot, we observe that, compared to the other vectors, both the $p$ and $p_{\text{random}}$ vectors display clustering patterns. This suggests that the $p$ and $p_{\text{random}}$ vectors extracted from different videos are highly similar, and the visual and temporal additional information in AI-generated videos shows significant clustering.

\section{Related Work}
\label{sec:6}
%In this section, we summarize related work on bias in information retrieval and AI-generated content detection.
%\subsection{Bias in Information Retrieval}
%\label{sec:6.1}

Bias in information retrieval has attracted significant attention. In \cite{mowshowitz2002assessing}, it first introduced the concept of bias, defining it, analyzing its sources, and proposing methods for evaluating bias, such as comparing search engine performance. Later studies focused on real-time methods for measuring bias in web search engines~\cite{mowshowitz2005measuring}. Research has since progressed in three main areas: analyzing bias sources in specific domains, exploring methods for assessing and mitigating bias, and investigating retrieval bias induced by AIGC-generated content.

% Research on assessing and mitigating bias has revealed a significant negative correlation between retrieval bias and performance~\cite{wilkie2014retrievability}. Subsequent studies addressed fairness across various systems and proposed mitigation strategies~\cite{yao2017new, yang2017measuring, geyik2019fairness}. Efforts to quantify bias in search engines, social networks, and recommendation services, along with fairness in collaborative filtering systems, were also explored~\cite{pitoura2018measuring}. Additional work focused on tackling fairness challenges in search and information retrieval systems~\cite{gao2021addressing, gao2021toward}.

Regarding methods for assessing and mitigating bias, research has explored the relationship between retrieval bias and performance, revealing a significant negative correlation~\cite{wilkie2014retrievability}. Later studies expanded on fairness in various systems and proposed mitigation strategies~\cite{yao2017new, yang2017measuring, geyik2019fairness}. Fairness in collaborative filtering recommendation systems and ranked outputs are examined, with efforts to quantify bias in search engines, social networks, and recommendation services~\cite{pitoura2018measuring}. Other studies concentrated on addressing fairness challenges in search and retrieval systems~\cite{gao2021addressing, gao2021toward}.

The rise of AIGC has introduced new challenges in retrieval bias. Research has explored the bias introduced by large language models (LLMs) in retrieval systems, revealing that neural retrieval models tend to prioritize AIGC-generated documents, a phenomenon known as source bias~\cite{dai2023llms}. Studies also show that objects in images generated by large vision-language models (LVLMs) exhibit more hallucination features compared to natural images~\cite{gao2024aigcs}. Additionally, it has been highlighted that synthetic images can introduce biases, with strategies proposed to mitigate these effects~\cite{xu2024invisible}. While prior research has not addressed source bias in video generation, this work validates its existence, identifies its origins in visual and temporal factors, and proposes solutions to mitigate the bias.

\section{Conclusion}

This study investigates the impact of AI-generated videos on text-video retrieval. We construct a comprehensive retrieval scenario that includes both real and AI-generated videos and conduct experiments based on this benchmark. The results show that AI-generated videos are preferentially retrieved by the model, appearing at the top of the retrieval list. As the proportion of AI-generated videos in the training set increases, the source bias becomes more pronounced. 
We analyze the reasons for the source bias. It not only originates from the visual information in AI-generated videos but also temporal information.
Finally, we employ a contrastive learning-based debiasing approach to alleviate the source bias and find that the additional information encoded by the generative model contributes to this bias.
The findings highlight the potential impact of AI-generated videos on text-video retrieval and offer valuable insights for future research.

\balance
\bibliographystyle{ACM-Reference-Format}
\bibliography{sample-base}

\newpage
\appendix
\onecolumn

% \chapter*{Appendix}
\setcounter{table}{0}
\section{Statistical Significance Analysis through Paired Hypothesis Testing}
\label{app:significance}

We performed rigorous statistical evaluation employing comprehensive paired T-tests to assess significance across our experimental framework. The analysis systematically examined:

\begin{itemize}
    \item All four video generation methodologies under investigation.
    \item Each of the three retrieval models implemented.
    \item Every dataset included in our study.
\end{itemize}

The statistical analysis yielded consistent and robust results, with all comparisons demonstrating significance at the conventional threshold (p < 0.05). Our experimental design treated the following as independent comparison groups for hypothesis testing:

\begin{itemize}
\item Group A: Text-real video pairs (ground truth)

\item Group B: Text-AI-generated video pairs (experimental condition)
\end{itemize}

This methodological approach enabled quantitative evaluation of similarity score distributions between authentic and synthesized video-text pairings across all experimental conditions. The statistically significant outcomes validate the reliability of our comparative findings.

\section{Limitation: Commercial Video Generation Models}

In this study, we relied on two open-source video generators, CogVideoX and OpenSora, which may limit the generalizability of our findings to commercial models (e.g., Sora, Pika) that exhibit superior temporal fidelity. However, we argue that the core conclusions of this paper remain valid even when considering such commercial models, and the "Visual-Temporal Induced Source Bias" phenomenon might be further amplified. 

Technically, both open-source and commercial video generation models embed additional visual-temporal information during the generation process, which can induce retrieval bias. While commercial models like Sora may demonstrate superior performance in temporal and visual aspects, this could exacerbate the observed "Visual-Temporal Induced Source Bias," as more natural temporal and visual information would be further recognized by retrieval models as "high-relevance features". To provide additional validation against closed-source generators, we performed retrieval experiments on 100 videos synthesized using PIKA, leveraging the InternVideo model. As shown in Table \ref{tab:pika_results}, the results are consistent with our previous conclusions, further supporting the robustness of our findings.

\begin{table}[htbp]
    \centering
    \caption{Retrieval results on 100 videos synthesized using PIKA with the InternVideo model.}
    \begin{tabular}{ccc}
        \toprule
        PIKA100 Metric & Relative$\Delta$ & Normalized$\Delta$ \\
        \midrule
        R@1 & -25.35 & -32.76 \\
        MeanR & -41.07 & -3.92 \\
        MixR & -22.14 & -12.23 \\
        \bottomrule
    \end{tabular}
    \label{tab:pika_results}
\end{table}

\section{Limitation: Video Generation Parameters}

Regarding FPS, changing it normally only affects frame sampling during video encoding, rather than fundamentally altering bias formation—this only slightly impacts test results.

For resolution changes, we used the moviepy library to reduce the resolution of OpenSora TextCond videos. Severe resolution reduction to 50×50 or 100×100 caused significant quality degradation and bias attenuation. In contrast, moderate downsampling to 424×240 preserved bias characteristics, consistent with our earlier results (as seen in the OpenSora VideoExt dataset in Table \ref{tab:resolution_impact}, which has a resolution of 424×240 and maintains bias characteristics), as verified through InternVideo-based retrieval analysis.

\begin{table}[htbp]
    \centering
    \caption{Impact of different resolutions on retrieval metrics}
    \begin{tabular}{ccccc}
        \toprule
        Resolution & R@1 & R@5 & MedR & MixR \\
        \midrule
        50×50 & 46.97 & 4.54 & 22.22 & 11.24 \\
        100×100 & 19.68 & 0.26 & -13.33 & -8.72 \\
        424×240 & -42.70 & -14.80 & -50.00 & -34.60 \\
        \bottomrule
    \end{tabular}
    \label{tab:resolution_impact}
\end{table}

\section{Future Work}

Our ongoing research will delve deeper into the fundamental causes of bias in AI-generated video content. While our current analysis has examined biases from visual, temporal, and internal vector representation perspectives, the precise underlying mechanisms warrant further investigation.

In subsequent studies, we intend to:

\begin{itemize}
    \item Employ more sophisticated video retrieval and generation models in our experimental framework.
    \item Conduct comprehensive analyses on the generalizability of existing debiasing approaches.
    \item Implement systematic keyword replacement in generated videos to quantitatively assess its impact on retrieval performance, thereby elucidating the relationship between keyword overemphasis and bias formation.
\end{itemize}

% \newpage

\section{CogVideoX TextCond and OpenSora TextCond}

\FloatBarrier
\begin{table*}[ht]

\caption{The retrieval performance of different models is evaluated on CogVideoX TextCond and OpenSora TextCond. When $Relative\Delta>0$ or $Normalized\Delta>0$, it indicates that the retrieval model tends to rank real videos higher. Conversely, \textcolor{red}{$Relative\Delta<0$} or \textcolor{red}{$Normalized\Delta<0$} suggests that the model tends to rank AI-generated videos higher. The absolute values of these metrics reflect the magnitude of the bias. $\text{Normalized}\Delta$ incorporates a penalty term to the original $Relative\Delta$, providing a more precise measurement of the bias.}
  \label{tab:main-exp-text-appendix}
  \centering
  % \setlength{\tabcolsep}{0.8mm}
  % \renewcommand{\arraystretch}{0.9}
  % \small
  \resizebox{0.85\textwidth}{!}{%
  \begin{tabular}{llcccccccccccc}
  \toprule
    \multicolumn{2}{l}{\makecell[l]{\hspace{1.5em}\textbf{Dataset}}}  & 
    \multicolumn{6}{c}{\makecell[l]{\hspace{3em}\textbf{CogVideoX TextCond}}} &
    \multicolumn{6}{c}{\makecell[l]{\hspace{3em}\textbf{OpenSora TextCond}}}
    \\
    % \midrule
     Model &   Metric & R@1 & R@5 & R@10 & MedR & MeanR & MixR & R@1 & R@5 & R@10 & MedR & MeanR & MixR \\
    \midrule
     \multirow{6}{*}{\makecell[c]{Alpro}}& REAL & 24.10  & 45.10  & 55.50  & 8.00  & 49.61 & - & 24.10  & 45.10  & 55.50  & 8.00  & 49.61 & -  \\
      & AI & 30.50  & 51.70  & 61.90  & 5.00  & 40.14 & - & 37.00  & 59.30  & 68.90  & 3.00  & 27.72 & -  \\
     & mixed-REAL & 10.10  & 34.60  & 45.50  & 14.00  & 82.94  & -  & 10.80  & 35.40  & 46.80  & 13.50  & 83.72  & -  \\ 
     & mixed-AI & 22.60  & 42.70  & 50.70  & 10.00  & 101.16  & -  & 24.50  & 49.50  & 56.10  & 6.00  & 69.39  & -  \\ 
     & $\cellcolor{gray!15}\text{Relative}\Delta$ & \cellcolor{gray!15}\textcolor{red}{-76.45}  & \cellcolor{gray!15}\textcolor{red}{-20.96}  & \cellcolor{gray!15}\textcolor{red}{-10.81}  & \cellcolor{gray!15}\textcolor{red}{-33.33}  & \cellcolor{gray!15}19.80  & \cellcolor{gray!15}\textcolor{red}{-29.99}  & \cellcolor{gray!15}\textcolor{red}{-77.62}  & \cellcolor{gray!15}\textcolor{red}{-33.22}  & \cellcolor{gray!15}\textcolor{red}{-18.08}  & \cellcolor{gray!15}\textcolor{red}{-76.92}  & \cellcolor{gray!15}\textcolor{red}{-18.71}  & \cellcolor{gray!15}\textcolor{red}{-57.75}  \\ 
    & $\cellcolor{gray!15}\text{Normalized}\Delta$ & \cellcolor{gray!15}\textcolor{red}{-53.01}  & \cellcolor{gray!15}\textcolor{red}{-2.59}  & \cellcolor{gray!15}2.83  & \cellcolor{gray!15}14.67  & \cellcolor{gray!15}41.02  & \cellcolor{gray!15}0.89  & \cellcolor{gray!15}\textcolor{red}{-35.39}  & \cellcolor{gray!15}3.05  & \cellcolor{gray!15}9.12  & \cellcolor{gray!15}18.32  & \cellcolor{gray!15}38.26  & \cellcolor{gray!15}7.06  \\ 
    % \hline
     \multirow{6}{*}{\makecell[c]{Frozen}} & REAL & 22.90  & 43.20  & 53.60  & 8.00  & 49.81 & - & 22.90  & 43.20  & 53.60  & 8.00  & 49.81 & -  \\
     & AI & 29.80  & 50.60  & 60.80  & 5.00  & 39.98 & - & 31.50  & 54.70  & 64.30  & 4.00  & 31.56 & -  \\
     & mixed-REAL &
    6.90  & 28.20  & 39.10  & 20.00  & 92.25  & -  & 8.90  & 31.40  & 41.40  & 17.00  & 90.35  & -  \\ 
     & mixed-AI &
    23.80  & 45.20  & 53.00  & 8.00  & 90.98  & -  & 25.50  & 46.80  & 55.40  & 7.00  & 72.41  & -  \\ 
     & $\cellcolor{gray!15}\text{Relative}\Delta$ &
    \cellcolor{gray!15}\textcolor{red}{-110.10}  & \cellcolor{gray!15}\textcolor{red}{-46.32}  & \cellcolor{gray!15}\textcolor{red}{-30.18}  & \cellcolor{gray!15}\textcolor{red}{-85.71}  & \cellcolor{gray!15}\textcolor{red}{-1.39}  & \cellcolor{gray!15}\textcolor{red}{-65.73}  & \cellcolor{gray!15}\textcolor{red}{-96.51}  & \cellcolor{gray!15}\textcolor{red}{-39.39}  & \cellcolor{gray!15}\textcolor{red}{-28.93}  & \cellcolor{gray!15}\textcolor{red}{-83.33}  & \cellcolor{gray!15}\textcolor{red}{-22.05}  & \cellcolor{gray!15}\textcolor{red}{-67.30}  \\ \multirow{-6}{*} & $\cellcolor{gray!15}\text{Normalized}\Delta$ &
    \cellcolor{gray!15}\textcolor{red}{-83.91}  & \cellcolor{gray!15}\textcolor{red}{-23.51}  & \cellcolor{gray!15}\textcolor{red}{-14.40}  & \cellcolor{gray!15}\textcolor{red}{-37.71}  & \cellcolor{gray!15}20.63  & \cellcolor{gray!15}\textcolor{red}{-33.66}  & \cellcolor{gray!15}\textcolor{red}{-64.89}  & \cellcolor{gray!15}\textcolor{red}{-10.89}  & \cellcolor{gray!15}\textcolor{red}{-5.44}  & \cellcolor{gray!15}\textcolor{red}{-13.76}  & \cellcolor{gray!15}23.08  & \cellcolor{gray!15}\textcolor{red}{-18.52}  \\ 
    % \hline
     \multirow{6}{*}{\makecell[c]{Intern\\Video}} & REAL & 40.60  & 66.70  & 75.20  & 2.00  & 22.27 & - & 40.60  & 66.70  & 75.20  & 2.00  & 22.27 & -  \\
     & AI & 40.20  & 64.00  & 73.40  & 2.00  & 25.30 & - & 47.20  & 71.50  & 78.40  & 2.00  & 17.85 & -  \\
     & mixed-REAL &
    19.60  & 52.30  & 63.50  & 5.00  & 43.39  & -  & 27.40  & 53.10  & 62.20  & 5.00  & 74.16  & -  \\ 
     & mixed-AI &
    27.60  & 56.10  & 64.90  & 4.00  & 56.31  & -  & 22.50  & 58.20  & 68.90  & 4.00  & 26.87  & -  \\ 
     & $\cellcolor{gray!15}\text{Relative}\Delta$ &
    \cellcolor{gray!15}\textcolor{red}{-33.90}  & \cellcolor{gray!15}\textcolor{red}{-7.01}  & \cellcolor{gray!15}\textcolor{red}{-2.18}  & \cellcolor{gray!15}\textcolor{red}{-22.22}  & \cellcolor{gray!15}25.92  & \cellcolor{gray!15}\textcolor{red}{-10.07}  & \cellcolor{gray!15}19.64  & \cellcolor{gray!15}\textcolor{red}{-9.16}  & \cellcolor{gray!15}\textcolor{red}{-10.22}  & \cellcolor{gray!15}\textcolor{red}{-22.22}  & \cellcolor{gray!15}\textcolor{red}{-93.61}  & \cellcolor{gray!15}\textcolor{red}{-32.06}  \\ 
     & $\cellcolor{gray!15}\text{Normalized}\Delta$ &
    \cellcolor{gray!15}\textcolor{red}{-34.89}  & \cellcolor{gray!15}\textcolor{red}{-11.17}  & \cellcolor{gray!15}\textcolor{red}{-6.31}  & \cellcolor{gray!15}\textcolor{red}{-22.22}  & \cellcolor{gray!15}13.06  & \cellcolor{gray!15}\textcolor{red}{-14.68}  & \cellcolor{gray!15}34.67  & \cellcolor{gray!15}\textcolor{red}{-1.66}  & \cellcolor{gray!15}\textcolor{red}{-3.27}  & \cellcolor{gray!15}\textcolor{red}{-22.22}  & \cellcolor{gray!15}\textcolor{red}{-71.32}  & \cellcolor{gray!15}\textcolor{red}{-19.62} 
    \\ \bottomrule 
  \end{tabular}
  }
\end{table*}
\FloatBarrier

\section{OpenSora ImageCond and OpenSora VideoExt}

\FloatBarrier
\begin{table*}[ht]
    \belowrulesep=0pt
    \aboverulesep=0pt
  \caption{The retrieval performance of different models is evaluated on the OpenSora ImageCond and OpenSora VideoExt datasets. For the definitions of $\text{Relative}\Delta$ and $Normalized\Delta$, please refer to Table \ref{tab:main-exp-text-appendix}.}
  \label{tab:main-exp-picture-appendix}
  \centering
  \setlength{\tabcolsep}{0.8mm}
  \renewcommand{\arraystretch}{0.85}
  \small
  % \rowcolors{2}{white}{gray!15}
  \resizebox{0.85\textwidth}{!}{%
  \begin{tabular}{llcccccccccccc}
  \toprule
    \multicolumn{2}{l}{\makecell[l]{\hspace{1.5em}\textbf{Dataset}}}  & 
    \multicolumn{6}{c}{\makecell[l]{\hspace{3em}\textbf{OpenSora ImageCond}}} &
    \multicolumn{6}{c}{\makecell[l]{\hspace{3em}\textbf{OpenSora VideoExt}}} 
    \\
     Model & Metric & R@1 & R@5 & R@10 & MedR & MeanR & MixR & R@1 & R@5 & R@10 & MedR & MeanR & MixR \\
    \midrule
     \multirow{6}{*}{\makecell[c]{Alpro}} & REAL & 24.1 & 45.1 & 55.5 & 8 & 49.61 & - & 24.1 & 45.1 & 55.5 & 8 & 49.61 & -  \\
      & AI & 29.6 & 54.5 & 63.2 & 4 & 33.59 & - & 32.1 & 54.5 & 65.5 & 4 & 36.42 & -  \\
     & mixed-REAL & 8 & 33.7 & 43.5 & 15.5 & 94.31 & - & 8.7 & 33.2 & 41.8 & 17 & 95.90 & -  \\
     & mixed-AI & 22.4 & 45.4 & 55.5 & 7 & 70.33 & - & 23.7 & 47.2 & 57.6 & 7 & 75.38 & -  \\
     & $\cellcolor{gray!15}\text{Relative}\Delta$ & $\cellcolor{gray!15}\textcolor{red}{-94.74}$ & $\cellcolor{gray!15}\textcolor{red}{-29.58}$ & $\cellcolor{gray!15}\textcolor{red}{-24.24}$ & $\cellcolor{gray!15}\textcolor{red}{-75.56}$ & $\cellcolor{gray!15}\textcolor{red}{-29.13}$ & $\cellcolor{gray!15}\textcolor{red}{-66.48}$ & $\cellcolor{gray!15}\textcolor{red}{-92.59}$ & $\cellcolor{gray!15}\textcolor{red}{-34.83}$ & $\cellcolor{gray!15}\textcolor{red}{-31.79}$ & $\cellcolor{gray!15}\textcolor{red}{-83.33}$ & $\cellcolor{gray!15}\textcolor{red}{-23.97}$ & $\cellcolor{gray!15}\textcolor{red}{-66.63}$  \\
     & $\cellcolor{gray!15}\text{Normalized}\Delta$ &
    $\cellcolor{gray!15}\textcolor{red}{-74.26}$ & $\cellcolor{gray!15}\textcolor{red}{-9.35}$ & $\cellcolor{gray!15}\textcolor{red}{-5.36}$ & $\cellcolor{gray!15}\textcolor{red}{-5.99}$ & $\cellcolor{gray!15}9.61$ & $\cellcolor{gray!15}\textcolor{red}{-23.55}$ & $\cellcolor{gray!15}\textcolor{red}{-64.12}$ & $\cellcolor{gray!15}\textcolor{red}{-9.13}$ & $\cellcolor{gray!15}\textcolor{red}{-12.91}$ & $\cellcolor{gray!15}\textcolor{red}{-13.76}$ & $\cellcolor{gray!15}6.87$ & $\cellcolor{gray!15}\textcolor{red}{-23.67}$  \\
    % \hline
    \multirow{6}{*}{\makecell[c]{Frozen}} & REAL & 22.9 & 43.2 & 53.6 & 8 & 49.81 & - & 22.9 & 43.2 & 53.6 & 8 & 49.811 & -  \\
     & AI & 25.7 & 50.6 & 62.6 & 5 & 37.93 & - & 28.3 & 51 & 60.1 & 5 & 37.34 & -  \\
     & mixed-REAL &
    9.1 & 31 & 42.3 & 18 & 94.78 & - & 8.3 & 29.3 & 39.2 & 21 & 104.51 & -  \\
     & mixed-AI &
    18.9 & 40.3 & 51.7 & 9 & 80.01 & - & 21.6 & 45.5 & 53.6 & 8 & 71.89 & -  \\
     & $\cellcolor{gray!15}\text{Relative}\Delta$ &
   $\cellcolor{gray!15}\textcolor{red}{-70}$ & $\cellcolor{gray!15}\textcolor{red}{-26.09}$ & $\cellcolor{gray!15}\textcolor{red}{-20}$ & $\cellcolor{gray!15}\textcolor{red}{-66.67}$ & $\cellcolor{gray!15}\textcolor{red}{-16.9}$ & $\cellcolor{gray!15}\textcolor{red}{-51.19}$ & $\cellcolor{gray!15}\textcolor{red}{-88.96}$ & $\cellcolor{gray!15}\textcolor{red}{-43.32}$ & $\cellcolor{gray!15}\textcolor{red}{-31.03}$ & $\cellcolor{gray!15}\textcolor{red}{-89.66}$ & $\cellcolor{gray!15}\textcolor{red}{-36.99}$ & $\cellcolor{gray!15}\textcolor{red}{-71.87}$  \\ 
    & $\cellcolor{gray!15}\text{Normalized}\Delta$ &
    $\cellcolor{gray!15}\textcolor{red}{-58.48}$ & $\cellcolor{gray!15}\textcolor{red}{-13.34}$ & $\cellcolor{gray!15}\textcolor{red}{-4.22}$ & $\cellcolor{gray!15}\textcolor{red}{-18.67}$ & $\cellcolor{gray!15}10.34$ & $\cellcolor{gray!15}\textcolor{red}{-22.27}$ & $\cellcolor{gray!15}\textcolor{red}{-67.87}$ & $\cellcolor{gray!15}\textcolor{red}{-22.65}$ & $\cellcolor{gray!15}\textcolor{red}{-14.47}$ & $\cellcolor{gray!15}\textcolor{red}{-41.66}$ & $\cellcolor{gray!15}\textcolor{red}{-8.21}$ & $\cellcolor{gray!15}\textcolor{red}{-39.25}$  \\ 
    % \hline
    \multirow{6}{*}{\makecell[c]{Intern\\Video}} & REAL & 40.6 & 66.7 & 75.2 & 2 & 22.27 & - & 40.6 & 66.7 & 75.2 & 2 & 22.27 & -  \\
     & AI & 42.7 & 70.2 & 78.9 & 2 & 18.62 & - & 46.6 & 71 & 78.6 & 2 & 17.62 & -  \\
     & mixed-REAL &
    29.1 & 52.5 & 61.9 & 4 & 83.65 & - & 28.2 & 53.6 & 62.8 & 4 & 75.72 & -  \\
     & mixed-AI &
    16.2 & 56.3 & 70.3 & 4 & 26.31 & - & 20.4 & 56.6 & 68.8 & 4 & 26.57 & -  \\
     & $\cellcolor{gray!15}\text{Relative}\Delta$ &
    $\cellcolor{gray!15}56.95$ & $\cellcolor{gray!15}\textcolor{red}{-6.99}$ & $\cellcolor{gray!15}\textcolor{red}{-12.71}$ & $\cellcolor{gray!15}0.00$ & $\cellcolor{gray!15}\textcolor{red}{-104.29}$ & $\cellcolor{gray!15}\textcolor{red}{-15.78}$ & $\cellcolor{gray!15}32.1$ & $\cellcolor{gray!15}\textcolor{red}{-5.44}$ & $\cellcolor{gray!15}\textcolor{red}{-9.12}$ & $\cellcolor{gray!15}0.00$ & $\cellcolor{gray!15}\textcolor{red}{-96.08}$ & $\cellcolor{gray!15}\textcolor{red}{-21.33}$  \\
     & $\cellcolor{gray!15}\text{Normalized}\Delta$ &
    $\cellcolor{gray!15}61.99$ & $\cellcolor{gray!15}\textcolor{red}{-3.59}$ & $\cellcolor{gray!15}\textcolor{red}{-7.6}$ & $\cellcolor{gray!15}0.00$ & $\cellcolor{gray!15}\textcolor{red}{-86.22}$ & $\cellcolor{gray!15}\textcolor{red}{-8.08}$ & $\cellcolor{gray!15}45.86$ & $\cellcolor{gray!15}2.8$ & $\cellcolor{gray!15}\textcolor{red}{-2.87}$ & $\cellcolor{gray!15}0.00$ & $\cellcolor{gray!15}\textcolor{red}{-72.5}$ & $\cellcolor{gray!15}\textcolor{red}{-8.88}$  \\ 
    \bottomrule 
  \end{tabular}
  }
\end{table*}
\FloatBarrier

\newpage
\section{Shuffled OpenSora TextCond}

\FloatBarrier
\begin{table*}[ht]
  \belowrulesep=0pt
  \aboverulesep=0pt
   \caption{The retrieval performance of Text-Video Retrieval Models on the OpenSora TextCond dataset is assessed following the shuffling of video frame order. 'Random' denotes the shuffling of both real and AI-generated videos, while 'Random-only-AI' refers to shuffling only the AI-generated videos. For the definitions of $\text{Relative}\Delta$ and $Normalized\Delta$, please refer to Table~\ref{tab:main-exp-text-appendix}.}
  \label{tab:main-exp-random-appendix}
  \centering
  \setlength{\tabcolsep}{0.8mm}
  \renewcommand{\arraystretch}{0.9}
  \small
  % \rowcolors{2}{white}{gray!15}
  \resizebox{0.90\textwidth}{!}{%
  \begin{tabular}{llcccccccccccc}
  \toprule
    \multicolumn{2}{c}{\makecell[l]{\hspace{1.5em}\textbf{Dataset}}}  & 
    \multicolumn{6}{c}{\makecell[l]{\hspace{3em}\textbf{Random}}} &
    \multicolumn{6}{c}{\makecell[l]{\hspace{3em}\textbf{Random-only-AI}}} 
    
    \\
     Model & Metric & R@1 & R@5 & R@10 & MedR & MeanR & MixR & R@1 & R@5 & R@10 & MedR & MeanR & MixR \\
    \midrule
     \multirow{6}{*}{\makecell[c]{Alpro}} & REAL & 23.5 & 43.6 & 52.6 & 9 & 53.71 & - & 24.1 & 45.1 & 55.5 & 8 & 49.61 & - \\
      & AI & 37.0 & 59.5 & 68.5 & 3 & 28.56 & - & 37.0 & 59.5 & 68.5 & 3 & 28.56 & -  \\
     & mixed-REAL & 9.7 & 34.1 & 43.6 & 16 & 90.49 & - & 10.3 & 35.3 & 46.3 & 13 & 84.72 & - \\
     & mixed-AI & 25.8 & 49.2 & 56.9 & 6 & 71.15 & - & 25.9 & 48.9 & 57.4 & 6 & 69.77 & - \\
     & $\cellcolor{gray!15}\text{Relative}\Delta$ & $\cellcolor{gray!15}\textcolor{red}{-90.7}$ & $\cellcolor{gray!15}\textcolor{red}{-36.25}$ & $\cellcolor{gray!15}\textcolor{red}{-26.47}$ & $\cellcolor{gray!15}\textcolor{red}{-90.91}$ & $\cellcolor{gray!15}\textcolor{red}{-23.93}$ & $\cellcolor{gray!15}\textcolor{red}{-68.51}$ & $\cellcolor{gray!15}\textcolor{red}{-86.19}$ & $\cellcolor{gray!15}\textcolor{red}{-32.3}$ & $\cellcolor{gray!15}\textcolor{red}{-21.41}$ & $\cellcolor{gray!15}\textcolor{red}{-73.68}$ & $\cellcolor{gray!15}\textcolor{red}{-19.35}$ & $\cellcolor{gray!15}\textcolor{red}{-59.74}$ \\
     & $\cellcolor{gray!15}\text{Normalized}\Delta$ & $\cellcolor{gray!15}\textcolor{red}{-46.07}$ & $\cellcolor{gray!15}2.1$ & $\cellcolor{gray!15}4.37$ & $\cellcolor{gray!15}13.44$ & $\cellcolor{gray!15}37.58$ & $\cellcolor{gray!15}1.65$ & $\cellcolor{gray!15}\textcolor{red}{-43.96}$ & $\cellcolor{gray!15}3.68$ & $\cellcolor{gray!15}6.12$ & $\cellcolor{gray!15}21.56$ & $\cellcolor{gray!15}34.85$ & $\cellcolor{gray!15}4.15$ \\ 
     \multirow{6}{*}{\makecell[c]{Frozen}} & REAL & 20.5 & 41.9 & 51.8 & 9 & 56.19 & - & 22.9 & 43.2 & 53.6 & 8 & 49.81 & -  \\
     & AI & 30.7 & 54.9 & 65.4 & 4 & 32.57 & - & 30.7 & 54.9 & 65.4 & 4 & 32.57 & - \\
     & mixed-REAL &
    7.2 & 29.6 & 40.6 & 19 & 102.92 & - & 9.3 & 32.9 & 42.8 & 16 & 90.58 & - \\
     & mixed-AI &
    24.3 & 46 & 54.2 & 7 & 74.01 & - & 24 & 45.1 & 54.2 & 8 & 74.48 & -  \\
     & $\cellcolor{gray!15}\text{Relative}\Delta$ &
    $\cellcolor{gray!15}\textcolor{red}{-108.57}$ & $\cellcolor{gray!15}\textcolor{red}{-43.39}$ & $\cellcolor{gray!15}\textcolor{red}{-28.69}$ & $\cellcolor{gray!15}\textcolor{red}{-92.31}$ & $\cellcolor{gray!15}\textcolor{red}{-32.69}$ & $\cellcolor{gray!15}\textcolor{red}{-77.86}$ & $\cellcolor{gray!15}\textcolor{red}{-88.29}$ & $\cellcolor{gray!15}\textcolor{red}{-31.28}$ & $\cellcolor{gray!15}\textcolor{red}{-23.51}$ & $\cellcolor{gray!15}\textcolor{red}{-66.67}$ & $\cellcolor{gray!15}\textcolor{red}{-19.51}$ & $\cellcolor{gray!15}\textcolor{red}{-58.16}$  \\ 
    & $\cellcolor{gray!15}\text{Normalized}\Delta$ &
    $\cellcolor{gray!15}\textcolor{red}{-68.73}$ & $\cellcolor{gray!15}\textcolor{red}{-11.52}$ & $\cellcolor{gray!15}\textcolor{red}{-1.83}$ & $\cellcolor{gray!15}\textcolor{red}{-12.31}$ & $\cellcolor{gray!15}20.83$ & $\cellcolor{gray!15}\textcolor{red}{-20.07}$ & $\cellcolor{gray!15}\textcolor{red}{-59.19}$ & $\cellcolor{gray!15}\textcolor{red}{-4.67}$ & $\cellcolor{gray!15}0.34$ & $\cellcolor{gray!15}2.9$ & $\cellcolor{gray!15}22.6$ & $\cellcolor{gray!15}\textcolor{red}{-11.23}$  \\ 
    \multirow{6}{*}{\makecell[c]{Intern\\Video}} & REAL &
    39.6 & 65.1 & 73.2 & 2 & 24.56 & - & 40.6 & 66.7 & 75.20 & 2.00 & 22.27 & -  \\
     & AI & 47 & 69.9 & 77.9 & 2 & 18.04 & - & 47 & 69.9 & 77.9 & 2 & 18.04 & -  \\
     & mixed-REAL & 22.9 & 48.4 & 57.1 & 6 & 86.24 & - & 28.3 & 54.1 & 63 & 4 & 75.43 & -  \\
     & mixed-AI &
    28 & 61.1 & 71 & 3 & 25.37 & - & 20.8 & 56 & 68.4 & 4 & 27.36 & -  \\
     & $\cellcolor{gray!15}\text{Relative}\Delta$ &
    $\cellcolor{gray!15}\textcolor{red}{-20.04}$ & $\cellcolor{gray!15}\textcolor{red}{-23.2}$ & $\cellcolor{gray!15}\textcolor{red}{-21.7}$ & $\cellcolor{gray!15}\textcolor{red}{-66.67}$ & $\cellcolor{gray!15}\textcolor{red}{-109.07}$ & $\cellcolor{gray!15}\textcolor{red}{-65.26}$ & $\cellcolor{gray!15}30.55$ & $\cellcolor{gray!15}\textcolor{red}{-3.45}$ & $\cellcolor{gray!15}\textcolor{red}{-8.22}$ & $\cellcolor{gray!15}0$ & $\cellcolor{gray!15}\textcolor{red}{-93.53}$ & $\cellcolor{gray!15}\textcolor{red}{-20.99}$  \\
     & $\cellcolor{gray!15}\text{Normalized}\Delta$ &
    $\cellcolor{gray!15}\textcolor{red}{-2.95}$ & $\cellcolor{gray!15}\textcolor{red}{-12.29}$ & $\cellcolor{gray!15}\textcolor{red}{-14.59}$ & $\cellcolor{gray!15}\textcolor{red}{-66.67}$ & $\cellcolor{gray!15}\textcolor{red}{-78.14}$ & $\cellcolor{gray!15}\textcolor{red}{-49.25}$ & $\cellcolor{gray!15}45.16$ & $\cellcolor{gray!15}3.31$ & $\cellcolor{gray!15}\textcolor{red}{-3.53}$ & $\cellcolor{gray!15}0$ & $\cellcolor{gray!15}\textcolor{red}{-72.32}$ & $\cellcolor{gray!15}\textcolor{red}{-9.05}$  \\ 
    \bottomrule 
  \end{tabular}
  }
\end{table*}
\FloatBarrier
\end{document}